\documentclass[12pt,a4paper]{article}
\usepackage{amssymb,amsmath}
\usepackage[dvips]{lscape,graphicx}
\usepackage{color}
\usepackage{cite}
\usepackage{ulem}

\newcommand{\bc}{\begin{center}}
\newcommand{\ec}{\end{center}}
\newcommand{\bd}{\begin{displaymath}}
\newcommand{\ed}{\end{displaymath}}
\newcommand{\be}{\begin{equation}}
\newcommand{\ee}{\end{equation}}
\newcommand{\ba}{\begin{array}}
\newcommand{\ea}{\end{array}}
\newcommand{\bt}{\begin{tabular}}
\newcommand{\et}{\end{tabular}}

\newcommand{\ds}{\displaystyle}

\newcommand{\lsim}{\raisebox{-0.13cm}{~\shortstack{$<$ \\[-0.07cm] $\sim$}}~}
\newcommand{\gsim}{\raisebox{-0.13cm}{~\shortstack{$>$ \\[-0.07cm] $\sim$}}~}

\def\at{\alpha_t}
\def\ab{\alpha_b}
\def\as{\alpha_s}
\def\atau{\alpha_{\tau}}

\def\oatab{O(\at\ab)}
\def\oatas{O(\at\as)}
\def\oabas{O(\ab\as)}

\def\oatq{O(\at^2)}
\def\oabq{O(\ab^2)}
\def\oatauq{O(\atau^2)}
\def\oabatau{O(\ab \atau)}

\begin{document}

\title{
\vspace*{-3cm}
\phantom{h} \hfill\mbox{\small ADP-14-29-T887}\\[-1.7cm]
\phantom{h} \hfill\mbox{\small KA-TP-29-2014}\\[-1.7cm]
\phantom{h} \hfill\mbox{\small SFB/CPP-14-79}
\\[1.5cm]
\textbf{Non-Standard Higgs Decays\\ in $U(1)$
  Extensions of the MSSM\\[4mm]}}

\date{}
\author{
P.~Athron${}^{a,c}$,
M.~M\"{u}hlleitner$^{b}$,
R.~Nevzorov${}^{a}$\footnote{On leave of absence from the Theory Department, SSC RF ITEP of NRC ``Kurchatov Institute", Moscow, Russia.} \, and
A.G.~Williams${}^{a}$\\[9mm]
{\small\it $^a$ ARC Centre of Excellence for Particle Physics at the Terascale and CSSM,}\\
{\small\it School of Chemistry and Physics, University of Adelaide, Adelaide SA 5005, Australia}\\[3mm]
{\small\it $^b$ Institute for Theoretical Physics, Karlsruhe Institute of Technology,} \\
{\small\it 76128 Karlsruhe, Germany.}\\[3mm]
{\small\it $^c$ ARC Centre of Excellence for Particle Physics at the Terascale,}\\
{\small\it School of Physics, Monash University, Melbourne VIC 3800, Australia}\\
}

\maketitle

\begin{abstract}
\noindent
In $U(1)$ extensions of the Minimal Supersymmetric extension of the Standard
Model there is a simple mechanism that leads to a heavy $Z'$
boson with a mass which is substantially larger than the supersymmetry breaking
scale. This mechanism may also result in a pseudoscalar state that is
light enough for decays of the
  $125\,\mbox{GeV}$ Standard Model-like Higgs boson into a
pair of such pseudoscalars to be kinematically allowed.  We study
these decays within $E_6$ inspired supersymmetric models with an exact custodial
symmetry that forbids tree-level flavor-changing transitions and the
most dangerous baryon and lepton number violating operators. We argue
that the branching ratio of the lightest Higgs boson decays into a
pair of the light pseudoscalar states may not be negligibly small.
\end{abstract}
\thispagestyle{empty}
\vfill
\newpage
\setcounter{page}{1}

\section{Introduction}
Searches for physics beyond the Standard Model (SM) by the LHC
experiments ATLAS and CMS have set rather stringent constraints on the
masses of supersymmetric (SUSY) particles and of $Z'$ bosons. Indeed,
in the case of the $E_6$ inspired models the LHC data exclude $Z'$
resonances with masses $M_{Z'}$ below $2.5\,\mbox{TeV}$ \cite{1,2}. In
the simplest $U(1)$ extensions of the Minimal Supersymmetric extension
of the Standard Model (MSSM) the extra $U(1)$ gauge symmetry is
normally broken by the vacuum expectation value (VEV) of the scalar
component of a superfield $S$, which is a singlet under the SM gauge
group and carries a non-zero $U(1)$ charge\footnote{In the literature
  such states are often referred to
    as SM singlets and this convention will also be followed in this
    paper.}. Since the VEV $\langle S\rangle$ of $S$ and the mass $M_{Z'}$ of the
$Z'$ boson are determined by the SUSY breaking scale in these models,
the multi-TeV $Z'$ mass typically implies that other sparticles also
have multi-TeV masses. Such masses typically exceed even the limits on
the first and second generation squarks set by the LHC and are well above
the current sub–TeV limits on the third generation sfermions and on
additional Higgs bosons. It is therefore worthwhile considering
alternative realisations of $U(1)$ extensions, in which the $Z'$ boson
can be substantially heavier than the sparticles, and the
phenomenological implications of such mechanisms.

 Such scenarios may be realised when the extra $U(1)$ gauge symmetry,
 $U(1)^\prime$, is broken by two VEVs coming from the superfields $S$ and
 $\overline{S}$, which are both singlets under the SM gauge group but
 have opposite $U(1)^\prime$ charges.  In this case the $U(1)^\prime$
 $D$-term contribution to the scalar potential may force the minimum
 of this potential to be along the $D$-flat direction (see, for
 example, \cite{Kolda:1995iw}).  As a consequence the VEVs $\langle
 S\rangle$ and $\langle\overline{S}\rangle$ can be much larger than
 the SUSY breaking scale.

The simplest renormalisable superpotential of the SUSY model of the
type discussed above can be written as \be W_S=\sigma \phi S
\overline{S}\,,
\label{hd1}
\ee where $\phi$ is a scalar superfield that does not participate in
the gauge interactions. When the coupling $\sigma$ goes to zero the
corresponding tree-level scalar potential takes the form
\begin{equation}
V_S = m^2_S |S|^2 + m^2_{\overline{S}} |\overline{S}|^2 + m^2_{\phi} |\phi|^2
+\ds\frac{Q_S^2 g^{\prime \, 2}_1}{2}\left(|S|^2-|\overline{S}|^2\right)^2\,,
\label{hd2}
\end{equation}
where $m_S^2$, $m^2_{\overline{S}}$ and $m^2_{\phi}$ are soft SUSY
breaking mass parameters squared, while $g^\prime_1$ is the $U(1)'$ gauge
coupling and $Q_S$ is the $U(1)'$ charge of the SM singlet superfields
$S$ and $\overline{S}$. In Eq.~(\ref{hd2}) the last term is associated with the extra
$U(1)^\prime$ $D$-term contribution. In the limit $\langle S \rangle
= \langle \overline{S} \rangle$ this quartic term vanishes. If $(m^2_S
+ m^2_{\overline{S}})<0$ then there is a run--away direction in this
model, so that $\langle S \rangle = \langle \overline{S} \rangle \to
\infty$. When the $F$-terms from the interaction in
  the superpotential Eq.~(\ref{hd1}) are included this stabilizes the
  run-away direction and for small values  of the coupling $\sigma$ the
  SM singlet superfields tend to acquire large VEVs, i.e.
\be \langle
\phi \rangle \sim \langle S \rangle \simeq \langle \overline{S}
\rangle \sim \dfrac{1}{\sigma}\sqrt{|m^2_S + m^2_{\overline{S}}|}\,,
\label{hd3}
\ee
resulting in an extremely heavy $Z'$ boson.

Although the SUSY model mentioned above looks rather simple and
elegant it also possesses an additional accidental global $U(1)$
symmetry which can be associated with the Peccei-Quinn (PQ) symmetry
\cite{Peccei:1977hh}.  This symmetry is spontaneously broken by the
VEVs of the SM singlet superfields resulting in a massless axion
\cite{axion}. To avoid the appearance of this axion one needs to
include in the superpotential of Eq.~(\ref{hd1}) polynomial terms with
respect to the superfield $\phi$ which explicitly break the global
$U(1)$ symmetry. If the couplings that violate the PQ symmetry are very small,
then the particle spectrum of this SUSY model should contain a
pseudo-Goldstone boson which can be considerably lighter than all
sparticles and Higgs bosons. In fact, the corresponding pseudoscalar
Higgs state may be so light that the decay of the SM-like Higgs boson into a
pair of these states can be kinematically allowed.

In this article we consider such non-standard Higgs decays within well
motivated $E_6$ inspired extensions of the MSSM, with the particular
model described in section 2.  We focus on scenarios with an
approximate global $U(1)$ symmetry that leads to a pseudo-Goldstone
boson in the particle spectrum.
The pseudo-Goldstone state in these scenarios is mainly a linear
superposition of the imaginary parts of the scalar components of the
SM singlet superfields $\phi$, $S$ and $\overline{S}$. The SM-like
Higgs boson on the other hand is predominantly a linear superposition
of the neutral components of the Higgs doublets $H_u$ and $H_{d}$, so
that the coupling of the pseudo-Goldstone state to the SM-like Higgs boson
can be expected to be somewhat suppressed.  However it can still lead to a
non-negligible branching ratio of the lightest Higgs decays into a
pair of pseudo-Goldstone bosons.

In this context it is worth noting that the decay rate of the SM-like
Higgs state into a pair of pseudoscalars was intensively studied
within the simplest extension of the MSSM, i.e.~the Next-to-Minimal
Supersymmetric extension of the SM (NMSSM). For reviews of
non-standard Higgs boson decays see \cite{Chang:2008cw}
and for a more recent work see e.g.~\cite{King:2014xwa}.
The NMSSM superpotential is given by \cite{review-nmssm}:
\begin{equation}
W_{\rm NMSSM}=\lambda S(H_d H_u)+\dfrac{\kappa}{3} S^3 + W_{\rm MSSM}(\mu=0)\,,
\label{hd5}
\end{equation}
where $W_{\rm MSSM}(\mu=0)$ is the MSSM superpotential with the
bilinear mass $\mu$ set to zero, and $\lambda$
and $\kappa$ are new
NMSSM-specific couplings\footnote{One of the motivations of the NMSSM
  is the dynamic generation of the supersymmetric Higgs mass parameter
  $\mu$ through the coupling term $SH_dH_u$ when the singlet field $S$
  acquires a vacuum expectation value $\langle S \rangle$,
  i.e.~$\mu= \lambda \langle S \rangle$.}. In the limit where the cubic
coupling $\kappa= 0$ the Lagrangian of the NMSSM is invariant under
the transformations of the PQ symmetry which leads to the massless
axion when it is spontaneously broken by the VEV $\langle S\rangle$.
If $\kappa$ is rather small then the NMSSM particle spectrum involves
one light scalar state and one light pseudoscalar state.  In addition,
if $\kappa\to 0$ and the SUSY breaking scale is of the order of a
$\mbox{TeV}$, then the lightest SUSY particle (LSP) is the lightest
neutralino $\tilde{\chi}^0_1$ and is predominantly singlino.  In this
case the LSP couplings to the SM particles are quite small resulting
in a relatively small annihilation cross section for
$\tilde{\chi}^0_1\tilde{\chi}^0_1\to \mbox{SM particles}$, which gives
rise to a relic density that is typically much larger than its
measured value. As a consequence it seems to be rather problematic to
find phenomenologically viable scenarios with a light pseudoscalar in
the case of the NMSSM with approximate PQ symmetry.  Nevertheless a
sufficiently light pseudoscalar can always be obtained by tuning the
parameters of the NMSSM.

In contrast to the NMSSM the mass of the lightest neutralino in the
SUSY model considered here does not become small when the PQ symmetry
violating couplings vanish. Moreover, even when all PQ symmetry
violating couplings are negligibly small, the LSP can be
higgsino-like. This allows a reasonable value for the dark matter
density to be obtained if the LSP has a mass below $1\,\mbox{TeV}$ (see,
for example \cite{ArkaniHamed:2006mb}). Thus the approximate PQ
symmetry can lead to phenomenologically viable scenarios with a light
pseudoscalar in this model.

The purpose of this paper is to study the implications of a light pseudoscalar Higgs
state, which can appear in this phenomenologically interesting model,
for the decays of the SM-like Higgs boson. In particular, we
investigate what values of branching ratios can be expected for the
SM-like Higgs decays into a pair of light pseudoscalars, taking into
account the constraints arising from the model itself and the experimental
restrictions due to the LHC Higgs search data.

The paper is organised as follows. In section 2 we briefly review
$E_6$ inspired SUSY models with exact custodial $\tilde{Z}^{H}_2$
symmetry. In section 3 we study the breakdown of the gauge symmetry
and the implications for Higgs phenomenology.  In section 4 we
discuss a set of benchmark scenarios that lead to the decays of the
lightest Higgs boson into a pair of pseudoscalar states. Our results
are summarized in section 5. In Appendix A the spectrum of the neutralino
states is examined.

\section{$E_6$ inspired SUSY models with exact $\tilde{Z}^{H}_2$ symmetry}

In this section, we briefly review the $E_6$ inspired SUSY models with
exact custodial $\tilde{Z}^{H}_2$ symmetry \cite{nevzorov}, which we
then use to demonstrate how light pseudoscalar states can appear in SUSY
models with an extra $U(1)'$ gauge symmetry. We also consider what kind of Higgs
decay rates they can lead to.


The breakdown of the $E_6$ symmetry at high energies may lead to
models based on rank-5 gauge groups with an additional $U(1)'$ factor
in comparison to the SM. In this case the extra $U(1)'$ gauge symmetry is
a linear combination of $U(1)_{\chi}$ and $U(1)_{\psi}$,
\be
U(1)'=U(1)_{\chi}\cos\theta^\prime+U(1)_{\psi}\sin\theta^\prime\,,
\label{hd4}
\ee
which are defined by the breakdown of the exceptional Lie group
$E_6$ into $SO(10)$, $E_6\to SO(10)\times U(1)_{\psi}$, and the
subsequent breakdown of $SO(10)$ into $SU(5)$, $SO(10)\to SU(5)\times
U(1)_{\chi}$ (for a review see
Refs.~\cite{Langacker:2008yv,E6-review}).  With additional Abelian
gauge symmetries it is important to ensure the cancellation of anomalies. In
any model based on the subgroup of $E_6$ the anomalies are canceled
automatically if the low-energy spectrum involves complete
$27$-plets. Consequently, in $E_6$ inspired SUSY models the particle
spectrum is extended to fill out three complete 27-dimensional
representations of $E_6$. Each $27$--plet, referred to as $27_i$ with
$i=1,2,3$, contains one generation of ordinary matter, a SM singlet
field $S_i$, that carries non--zero $U(1)'$ charge, up- and down-type Higgs
doublets $H^{u}_{i}$ and $H^{d}_{i}$ and charged $\pm 1/3$ exotic
quarks $D_i$ and $\bar{D}_i$.

Different aspects of the phenomenology of the $E_6$ inspired SUSY
models have been extensively studied in the past
\cite{E6-review,E6}. A few years ago the Tevatron and early LHC $Z'$
mass limits in these models were discussed in
Ref.~\cite{Accomando:2010fz}.
Collider signatures associated with the exotic quarks and squarks
have been considered in \cite{Kang:2007ib}. Previously, the
implications of $E_6$ inspired SUSY models with an additional $U(1)'$
gauge symmetry have been studied for electroweak symmetry
breaking (EWSB)
\cite{Langacker:1998tc,Cvetic:1997ky,Suematsu:1994qm,Keith:1997zb,Daikoku:2000ep},
neutrino physics \cite{Kang:2004ix, Ma:1995xk}, leptogenesis
\cite{Hambye:2000bn,King:2008qb}, electroweak (EW) baryogenesis
\cite{baryogen}, the muon anomalous magnetic moment \cite{g-2}, the
electric dipole moment of the electron \cite{Suematsu:1997tv} and of the tau lepton
\cite{GutierrezRodriguez:2006hb}, for lepton flavor violating processes
like $\mu\to e\gamma$ \cite{Suematsu:1997qt} and for CP-violation in the
Higgs sector \cite{Ham:2008fx}. The neutralino sector in $E_6$
inspired SUSY models was analysed in
\cite{Keith:1997zb,Suematsu:1997tv,GutierrezRodriguez:2006hb,Suematsu:1997qt,Suematsu:1997au,Keith:1996fv,Hesselbach:2001ri,Gherghetta:1996yr,E6neutralino-higgs}.
Such models have also been proposed as the solution to the tachyon
problems of anomaly-mediated SUSY breaking, via $U(1)^\prime$
$D$-term contributions \cite{Asano:2008ju}, and have been used in
combination with a generation symmetry to construct a model explaining
the fermion mass hierarchy and mixing \cite{Stech:2008wd}. The Higgs
sector and the
theoretical upper bound on the lightest Higgs boson mass in the $E_6$
inspired SUSY models were examined in \cite{Daikoku:2000ep,E6neutralino-higgs, King:2005jy,King:2005my,Accomando:2006ga,E6-higgs}.

Here we focus on the $E_6$ inspired SUSY extension of the SM based on
the low-energy SM gauge group together with an extra $U(1)_{N}$ gauge
symmetry in which right-handed neutrinos do not participate in the
gauge interactions. This corresponds to $\theta^\prime=\arctan\sqrt{15}$ in
Eq.~(\ref{hd4}).  In this Exceptional Supersymmetric Standard Model
(E$_6$SSM) \cite{King:2005jy,King:2005my} right-handed
neutrinos may be superheavy, shedding light on the origin of the mass
hierarchy in the lepton sector and providing a mechanism for the
generation of the baryon asymmetry in the Universe via leptogenesis
\cite{Hambye:2000bn,King:2008qb}. $E_6$ inspired SUSY models with an
additional $U(1)_{N}$ gauge symmetry have been studied in a variety of
contexts. Thus they have been investigated in \cite{Ma:1995xk} in the
context of non-standard neutrino models with extra singlets, in
\cite{Suematsu:1997au} from the point of view of $Z-Z'$ mixing, in
Refs.~\cite{Keith:1997zb,Suematsu:1997au,Keith:1996fv} the neutralino
sector was explored, in \cite{Keith:1997zb,King:2007uj}
the renormalisation group (RG) flow of the couplings was examined, and in
\cite{Suematsu:1994qm,Keith:1997zb,Daikoku:2000ep} EWSB was
studied. The presence of a $Z'$ boson and of exotic quarks as predicted
by the E$_6$SSM provides spectacular new physics signals at the LHC,
which were analysed in
\cite{King:2005jy,King:2005my,Accomando:2006ga,Howl:2007zi}. The
existence of light exotic particles also leads
to the non-standard decays of the SM-like Higgs boson that were
discussed in detail in \cite{Nevzorov:2013tta,Hall:2010ix}.  Within
the constrained version of the E$_6$SSM the particle spectrum and
associated collider signatures were studied in \cite{cE6SSM} and the
degree of fine tuning has recently been examined in
\cite{Athron:2013ipa}. The threshold corrections to the running gauge
and Yukawa couplings in the E$_6$SSM and their numerical impact in the
cE$_6$SSM were studied in detail in Ref.~\cite{Athron:2012pw}. Alternative
boundary conditions that take account of $D$-terms from the other
$U(1)$ gauge symmetry broken at the GUT scale were considered in
\cite{Miller:2012vn}, using the first or second generation sfermion
masses to constrain the GUT scale parameters. The renormalisation of
the VEVs in the E$_6$SSM was considered in \cite{Sperling:2013eva}.

The presence of exotic matter in the E$_6$SSM generically leads to
non-diagonal flavor transitions and rapid proton decay. A set of
discrete symmetries can be imposed in order to suppress these processes
\cite{King:2005jy,King:2005my}.  In this article we study the
non-standard Higgs decays mentioned above within the $E_6$ inspired
SUSY models with the extra $U(1)_{N}$ factor in which a single
discrete $\tilde{Z}^{H}_2$ symmetry forbids tree-level
flavor-changing transitions and the most dangerous baryon and lepton
number violating operators \cite{nevzorov}.
These models imply that $E_6$ or its subgroup is broken down to
 $SU(3)_C\times SU(2)_L\times U(1)_Y\times U(1)_{\psi}\times
 U(1)_{\chi}$ near the Grand Unification Theory (GUT) scale, which we
 denote as $M_X$. Below this scale $M_X$ the particle content of these
 SUSY models involves three copies of $27$--plets and a set of $M_{l}$
 and $\overline{M}_l$ supermultiplets from the incomplete $27'_l$ and
 $\overline{27'}_l$ representations of $E_6$, where $l$ runs over the
 different multiplets that are summarized below in
   Table \ref{tab-hd1}. All components of the complete $27_i$-plets
   are odd under the discrete symmetry $\tilde{Z}^{H}_2$, while
 the supermultiplets $\overline{M}_l$ can be either odd or even. The
 supermultiplets $M_{l}$ are even under the $\tilde{Z}^{H}_2$
 symmetry, and as a consequence they can be used for the breakdown of the gauge
 symmetry, while preserving the discrete symmetry. To ensure that the
 $SU(2)_L\times U(1)_Y\times U(1)_{\psi}\times U(1)_{\chi}$ symmetry
 is broken down to the $U(1)_{\rm em}$ gauge group associated with
 electromagnetism, these supermultiplets $M_{l}$ should
 contain the supermultiplets $H_u$, $H_d$, $S$ and a supermultiplet called
 $N^c_H$, which has the same quantum numbers as the right-handed
 neutrino. Just below the GUT scale the $U(1)_{\psi}\times
U(1)_{\chi}$ gauge symmetry is expected to be broken by the VEVs of
$N^c_H$ and $\overline{N}_H^c$ down to the $U(1)_{N}\times Z_{2}^{M}$
in these $E_6$ inspired models, where $Z_{2}^{M}=(-1)^{3(B-L)}$ is a
matter parity. This is possible because matter parity is a discrete
subgroup of $U(1)_{\psi}$ and $U(1)_{\chi}$. With such a breakdown
into $U(1)_{N}\times Z_{2}^{M}$ the right-handed neutrino mass can be
generated without breaking the remaining gauge symmetry and all exotic
states which originate from
 the $27_i$ representations of $E_6$ as well as ordinary quark and lepton
 states survive down to low energies. In general the large VEVs
 $\langle N^c_H \rangle \sim \langle \overline{N}_H^c \rangle \lesssim
 M_X$ also induce the large Majorana masses for the right-handed neutrinos
 allowing them to be used for the see-saw mechanism.  Since $N^c_H$
 and $\overline{N}_H^c$ acquire VEVs both of these supermultiplets must be even
 under the imposed $\tilde{Z}^{H}_2$ symmetry.

At the TeV scale the scalar components of the
superfields $H_u$, $H_d$ and $S$ play the role of Higgs fields.
The VEVs of the neutral scalar components
break the $SU(2)_L\times U(1)_Y\times U(1)_{N}$ gauge symmetry down to
$U(1)_{\rm em}$. Because of this the supermultiplets $H_u$, $H_d$ and
$S$ must also be even under the $\tilde{Z}^{H}_2$ symmetry.  In contrast, in the
simplest scenario $\overline{H}_u$ and $\overline{H}_d$ are expected
to be odd under this custodial symmetry so that they can combine with
the superposition of the corresponding components from the $27_i$,
forming vectorlike states that gain masses of order $M_X$.  The scalar
component of the superfield $\overline{S}$
may also acquire a non-zero VEV breaking
the $U(1)_{N}$ symmetry. If this is the case then $\overline{S}$ has
to be even under the $\tilde{Z}^{H}_2$ symmetry. When $\overline{S}$
is odd under the $\tilde{Z}^{H}_2$ symmetry then it can get combined
with the superposition of the appropriate components of $27_i$
resulting in the formation of superheavy vectorlike states with masses
$\sim M_X$.  The custodial $\tilde{Z}^{H}_2$ symmetry allows Yukawa
interactions in the superpotential that originate from $27'_l \times
27'_m \times 27'_n$ and $27'_l \times 27_i \times 27_k$ ($i,k=1,2,3$
and $l,m,n$ running over the multiplets given in
Table~\ref{tab-hd1}). It is easy to check that the
corresponding set of operators does not contain any
operators that lead to rapid proton decay. Since the set of supermultiplets
$M_{l}$ contains only one pair of doublets, $H_d$ and $H_u$, the
down-type quarks and charged leptons couple to just one Higgs doublet,
$H_d$, and the up-type quarks couple to $H_u$ only. As a result
flavor-changing processes are forbidden at tree-level.

Nonetheless, if the set of $\tilde{Z}^{H}_2$-even supermultiplets
$M_{l}$ involves only $H_u$, $H_d$, $S$ and $N^c_H$ then the Lagrangian
of the model is invariant not only with respect to the $U(1)_B$
associated with baryon number conservation but also
under $U(1)_D$ symmetry transformations \be D\to e^{i\alpha}
D\,,\qquad\qquad \overline{D}\to e^{-i\alpha}\overline{D}\,.
\label{hd6}
\ee
The $U(1)_D$ symmetry forbids renormalisable interactions through
which the exotic quarks $D$ can decay, thereby
ensuring that the lightest
exotic quark is very long-lived.  Indeed, as for $U(1)_B$ we expect
the $U(1)_D$ global symmetry to be broken by a set of
non-renormalisable operators that are suppressed by inverse powers of
$M_X$ or the Planck scale $M_{Pl}$. While these operators allow the lightest exotic
quark to decay, its lifetime tends to be considerably larger than the
age of the Universe.  Long-lived exotic quarks would have been
produced during the very early epochs of the Big Bang and those that
survive annihilation would subsequently have been confined in heavy
hadrons forming nuclear isotopes that would be present in terrestrial
matter. Various theoretical estimates \cite{43} show that if such
stable relics in the mass range from $1\,\mbox{GeV}$ to
$10\,\mbox{TeV}$ would exist in nature, today their concentration would
be ${\cal O}(10^{-10})$ per nucleon.  At the same time different
experiments set strong upper limits on the relative concentrations of
such nuclear isotopes from $10^{-15}$ to $10^{-30}$ per nucleon
\cite{42}. Therefore $E_6$ inspired models with very long-lived
exotic quarks are ruled out.

To ensure that the lightest exotic quarks decay within a reasonable
time in the simplest scenario, we supplement the set of
$\tilde{Z}^{H}_2$ even supermultiplets $M_{l}$ with $L_4$, where $L_4$
and $\overline{L}_4$ are lepton $SU(2)_L$ doublet supermultiplets that
originate from a pair of additional $27'_{L}$ and
$\overline{27'}_L$. The supermultiplets $L_4$ and $\overline{L}_4$
should form TeV scale vectorlike states to break the $U(1)_D$ symmetry
and render the lightest exotic quark unstable\footnote{The appropriate
  mass term $\mu_L L_4\overline{L}_4$ in the superpotential can be
  induced within SUGRA models just after the breakdown of local SUSY
  if the K\"ahler potential contains an extra term $(Z_L
  (L_4\overline{L}_4)+h.c)$ \cite{45}.}.  Therefore $L_4$ and
$\overline{L}_4$ both have to be even under the $\tilde{Z}^{H}_2$
symmetry.  In this case the baryon and lepton number conservation
implies that the exotic quarks are leptoquarks.

Here we assume that, in addition to $H_u$, $H_d$, $S$, $L_4$,
$\overline{L}_4$ $N^c_H$ and $\overline{N}_H^c$, the particle spectrum
below the scale $M_X$ involves $\tilde{Z}^{H}_2$-even superfields
$\overline{S}$ and $\phi$. The superfield $\phi$ does not
participate in the $SU(3)_C\times SU(2)_L\times U(1)_Y\times
U(1)_{\psi}\times U(1)_{\chi}$ gauge interactions but
  its scalar component acquires a non-zero VEV. Taking
into account that the components of the
superfields $\overline{S}$ and $\phi$ are expected to gain
$\mbox{TeV}$ scale masses whereas the right-handed neutrino
superfields are superheavy, the low-energy matter content in the $E_6$
inspired SUSY models discussed above involves
\be
\ba{c}
(Q_i,\,u^c_i,\,d^c_i,\,L_i,\,e^c_i)
+(D_i,\,\bar{D}_i)+(S_{i})+(H^u_{\alpha})+(H^d_{\alpha})\\[2mm]
+L_4+\overline{L}_4+S+\overline{S}+H_u+H_d+\phi\,,
\ea
\label{hd7}
\ee
where $\alpha=1,2$ runs over the first two generations and
$i=1,2,3$ runs over all three. We have denoted here the left-handed
quark and lepton doublets by $Q_i$ and $L_i$, respectively, and the
right-handed up- and down-type quarks and charged leptons by $u_i^c,
d_i^c$ and $e_i^c$. Neglecting all suppressed
non-renormalisable interactions, the low-energy effective
superpotential of these models can be written as \be \ba{rcl} W &=&
\lambda S (H_u H_d) - \sigma \phi S \overline{S} +
\dfrac{\kappa}{3}\phi^3+\dfrac{\mu}{2}\phi^2+\Lambda\phi\\[2mm] &&+
\lambda_{\alpha\beta} S (H^d_{\alpha} H^u_{\beta})+ \kappa_{ij} S
(D_{i} \overline{D}_{j}) + \tilde{f}_{i\alpha} S_{i} (H^d_{\alpha}
H_u) + f_{i\alpha} S_{i} (H_d H^u_{\alpha}) \\[2mm] &&+ g^D_{ij} (Q_i
L_4) \overline{D}_j+ h^E_{i\alpha} e^c_{i} (H^d_{\alpha} L_4) + \mu_L
L_4\overline{L}_4+ \tilde{\sigma} \phi L_4\overline{L}_4+ W_{\rm
  MSSM}(\mu=0)\,,  \ea
\label{hd8}
\ee
in terms of the dimensionless couplings
$\lambda,\sigma,\kappa,\lambda_{\alpha\beta}, \kappa_{ij},
\tilde{f}_{i\alpha}, f_{i\alpha}, g_{ij}^D,h_{i\alpha}^E,
\tilde{\sigma}$ and the dimensionful couplings $\mu,\mu_L$ and
$\Lambda$, with $i,j=1,2,3$ and $\alpha,\beta= 1,2$.

The gauge group and field content of the $E_6$ inspired SUSY models
under consideration can originate from the 5D and 6D orbifold GUT
models in which the splitting of GUT multiplets can be naturally
achieved \cite{nevzorov}. In these orbifold GUT models all GUT
relations between the Yukawa couplings can get spoiled while the
approximate unification of the SM gauge couplings still takes
place. From Eq.~(\ref{hd7}) it follows that extra matter beyond the
MSSM fills in complete $SU(5)$ representations in these models. As a
consequence the gauge coupling unification remains almost exact in the
one-loop approximation. It was also shown that in the two-loop
approximation the unification of the gauge couplings in the SUSY
models under consideration can be achieved for any phenomenologically
acceptable value of the strong coupling $\alpha_3(M_Z)$ at the scale
$M_Z$, consistent with the measured central low-energy value
\cite{King:2007uj,nevzorov}.

\begin{table}[ht]
\centering
\begin{tabular}{|c|c|c|c|c|c|c|c|c|c|}
\hline
                   &  $27_i$          &   $27_i$              &$27'_{H_u}$&$27'_{S}$&
$\overline{27'}_{H_u}$&$\overline{27'}_{S}$&$27'_N$&$27'_{L}$&$1$\\
& & &$(27'_{H_d})$& &$(\overline{27'}_{H_d})$& &$(\overline{27'}_N)$&$(\overline{27'}_L)$&\\
\hline
                   &$Q_i,u^c_i,d^c_i,$&$\overline{D}_i,D_i,$  & $H_u$     & $S$     &
$\overline{H}_u$&$\overline{S}$&$N^c_H$&$L_4$&$\phi$\\
                   &$L_i,e^c_i,N^c_i$ &  $H^d_{i},H^u_{i},S_i$& $(H_d)$   &         &
$(\overline{H}_d)$&&$(\overline{N}_H^c)$&$(\overline{L}_4)$&\\
\hline
$\tilde{Z}^{H}_2$  & $-$              & $-$                   & $+$       & $+$     &
$-$&$\pm$&$+$&$+$&$+$\\
\hline
$Z_{2}^{M}$        & $-$              & $+$                   & $+$       & $+$     &
$+$&$+$&$-$&$-$&$+$\\
\hline
$Z_{2}^{E}$        & $+$              & $-$                   & $+$       & $+$     &
$-$&$\pm$&$-$&$-$&$+$\\
\hline
\end{tabular}
\caption{Transformation properties of different components of $E_6$
  multiplets under the discrete symmetries $\tilde{Z}^H_2$,
  $Z_{2}^{M}$ and $Z_{2}^{E}$. A `$+$' denotes that the field is even under
  the discrete symmetry, a `$-$' means that the field is odd, while
  `$\pm$' denotes that the field may be even or odd depending on which
  construction is considered.  }
\label{tab-hd1}
\end{table}

For the analysis of the phenomenological implications of the SUSY
models discussed above it is convenient to introduce the $Z_{2}^{E}$
symmetry, which is defined such that $\tilde{Z}^{H}_2 =
Z_{2}^{M}\times Z_{2}^{E}$. The transformation properties of different
components of the $27_i$, $27'_l$ and $\overline{27'}_l$
supermultiplets under the $\tilde{Z}^{H}_2$, $Z_{2}^{M}$ and
$Z_{2}^{E}$ symmetries are summarized in Table~\ref{tab-hd1}. Since
the low-energy effective Lagrangian of the $E_6$ inspired SUSY models
studied here is invariant under the transformations of the $Z_{2}^{M}$
and $\tilde{Z}^{H}_2$ symmetries, the $Z_{2}^{E}$ symmetry associated
with the exotic states is also conserved. The invariance of the
Lagrangian under the $Z_{2}^{E}$ symmetry implies that in collider
experiments the exotic particles, which are odd under this symmetry,
can only be created in pairs and the lightest exotic state should be
absolutely stable. Using the method proposed in
\cite{Hesselbach:2007te} it was argued that the masses of the lightest
inert neutralino states\footnote{We use the terminology ``inert
  Higgs'' to denote the $H_\alpha^u, H_\alpha^d$ and
    $S$, whose scalar components do not develop VEVs. The
  fermionic components of these supermultiplets form inert neutralino
  and chargino states.}, which are predominantly
linear superpositions of the fermion components of the superfields
$S_i$ from complete $27_i$ representations of $E_6$, do not exceed
$60-65\,\mbox{GeV}$ \cite{Hall:2010ix}. Because of this the
corresponding states tend to be the lightest exotic particles in the spectrum.

The presence of lightest exotic particles with masses below
$60\,\mbox{GeV}$ gives rise to new decay channels of the SM-like Higgs
boson. Moreover, if these states are heavier than $5-10\;\mbox{GeV}$
({\it i.e.}~approximately above the bottom quark pair threshold)
then the SM-like Higgs state decays predominantly into the lightest
inert neutralinos while the total branching ratio into SM particles
gets strongly suppressed. Nowadays such scenarios are basically ruled
out. On the other hand if the lightest exotic particles have masses
below $5\,\mbox{GeV}$ their couplings to the SM-like
  Higgs state are small so that problems with non-standard Higgs
  decays can be avoided. However, as their couplings to the
  gauge bosons, quarks and leptons are also very small this results in a
cold dark matter density that is much larger than its measured value
because the corresponding annihilation cross section tends to be small.
The simplest phenomenologically viable scenarios imply that the
lightest inert neutralinos are substantially lighter than
$1\,\mbox{eV}$\footnote{The presence of very light neutral fermions in
  the particle spectrum might have interesting implications for
  neutrino physics (see, for example \cite{Frere:1996gb}).}. This can
be achieved if $f_{\alpha\beta}\sim \tilde{f}_{\alpha\beta}\lesssim 10^{-6}$. In this
case the lightest exotic particles form hot dark matter (dark
radiation) in the Universe but give only a very minor contribution to
the dark matter density.

The $Z_{2}^{M}$ symmetry conservation ensures that $R$-parity is also
conserved in the SUSY models discussed above. As also the $Z_{2}^{E}$ symmetry
is conserved there are two states possible that can be stable. This is
either the lightest $R$-parity even exotic state or the lightest
$R$-parity odd state with $Z_2^E=+1$. In the $E_6$ inspired models
studied here the stable state tends to be the lightest ordinary
neutralino state ({\it i.e.}~the lightest neutralino state with
$Z_{2}^{E}=+1$). Like in the MSSM, this state may account for all or
for some of the observed cold dark matter density.

As mentioned before, in the simplest case the sector responsible for
the breakdown of the $SU(2)_L\times U(1)_Y\times U(1)_{N}$ gauge
symmetry involves $H_u$, $H_d$ and $S$. For this case the Higgs sector
of the $E_6$ inspired SUSY models with the extra $U(1)_{N}$ factor was
explored in \cite{King:2005jy}.  If CP-invariance is preserved then
the Higgs spectrum in these models contains three CP-even, one
CP-odd and two charged states.  The singlet dominated CP-even state
is always almost degenerate with the $Z'$ gauge boson. In contrast to
the MSSM, the lightest Higgs boson in these models can be heavier than
$110-120\,\mbox{GeV}$ even at tree-level. In the two-loop
approximation the lightest Higgs boson mass does not exceed
$150-155\,\mbox{GeV}$ \cite{King:2005jy}. Recently, the RG flow of the
Yukawa couplings and the theoretical upper bound on the lightest Higgs
boson mass in these models were analysed in the vicinity of the
quasi-fixed point \cite{Nevzorov:2013ixa} that appears as a result of
the intersection of the invariant and quasi-fixed lines
\cite{Nevzorov:2001vj}.  It was argued that near the quasi-fixed
point the upper bound on the mass of the SM-like Higgs boson is
rather close to $125\,\mbox{GeV}$ \cite{Nevzorov:2013ixa}.

The qualitative pattern of the Higgs spectrum in the $E_6$ inspired
SUSY models with the extra $U(1)_{N}$ gauge symmetry and minimal Higgs
sector is determined by the Yukawa coupling
$\lambda$. When $\lambda <
g'_1$, where $g'_1$ is the gauge coupling associated with the
$U(1)_{N}$ gauge symmetry, the singlet dominated CP-even state is
very heavy and decouples from the rest of the spectrum, which makes
the Higgs spectrum indistinguishable from the one in the MSSM. If
$\lambda\gtrsim g'_1$ the Higgs spectrum has an extremely hierarchical
structure, which is rather similar to the one that arises in the NMSSM
with the approximate PQ symmetry
\cite{Miller:2005qua,Nevzorov:2004ge}. As a consequence the
mass matrix of the CP--even Higgs sector can be diagonalised using a
perturbative expansion
\cite{Nevzorov:2004ge,Nevzorov:2001um}. In this case the mass of
the second lightest CP-even Higgs state is set by the $Z'$ boson
mass, while the heaviest CP-even, CP-odd and charged states are
almost degenerate and lie beyond the multi-TeV range.

\section{The Higgs Sector}
\label{Sec:Higgs-sector}
\subsection{The Higgs Potential and Gauge Symmetry Breaking}

As was mentioned in the previous section, the sector responsible for
breaking the gauge symmetry in the SUSY model under consideration
includes two Higgs doublets, $H_u$ and $H_d$, as well as the SM
singlet fields $S$, $\overline{S}$ and $\phi$.

The interactions between these fields are determined by the structure of the gauge
interactions and by the superpotential in Eq.~(\ref{hd8}). The
resulting Higgs potential reads
\be
\ba{rcl}
V&=&V_F+V_D+V_{\rm soft}+\Delta V\, ,\\[2mm]
V_F&=&\lambda^2|S|^2(|H_d|^2+|H_u|^2) + \left|\lambda (H_d H_u) - \sigma \phi \overline{S}\right|^2
+ \sigma^2 |\phi|^2 |S|^2 + \\[2mm]
& + &\left| - \sigma (S \overline{S}) + \kappa \phi^2 + \mu \phi + \Lambda \right|^2\,,\\[2mm]
V_D&=&\ds \sum_{a=1}^3 \frac{g_2^2}{8}\left(H_d^\dagger \sigma_a H_d+H_u^\dagger \sigma_a H_u\right)^2
+\frac{{g'}^2}{8}\left(|H_d|^2-|H_u|^2\right)^2+\\[2mm]
&&+\ds\frac{g^{\prime \,2}_1}{2}\left(\tilde{Q}_{H_d} |H_d|^2+\tilde{Q}_{H_u} |H_u|^2+\tilde{Q}_S |S|^2 - \tilde{Q}_S |\overline{S}|^2 \right)^2\,,\\[2mm]
V_{soft}&=&m_{S}^2 |S|^2 + m_{\overline{S}}^2 |\overline{S}|^2 + m_{H_d}^2|H_d|^2 + m_{H_u}^2|H_u|^2 +
m^2_{\phi}|\phi|^2+ \biggl[\lambda A_{\lambda} S (H_u H_d) \\[2mm]
&-& \sigma A_{\sigma} \phi (S \overline{S}) + \dfrac{\kappa}{3} A_{\kappa} \phi^3 + B\dfrac{\mu}{2} \phi^2 + \xi \Lambda \phi + h.c.\biggr]\,,
\ea
\label{hd9}
\ee where $H_d^T=(H_d^0,\,H_d^{-})$, $H_u^T=(H_u^{+},\,H_u^{0})$ and
$(H_d H_u)=H_u^{+}H_d^{-}-H_u^{0}H_d^{0}$, and $\tilde{Q}_{H_d}$,
$\tilde{Q}_{H_u}$ and $\tilde{Q}_S$ are the effective $U(1)_N$ charges
of $H_d$, $H_u$ and $S$. Furthermore $\sigma_a$
  ($a=1,2,3$) denote the three Pauli matrices.  At tree-level the Higgs potential in
Eq.~(\ref{hd9}) is described by the sum of the first three
terms. $V_F$ and $V_D$ contain the $F$-and $D$-term contributions
that do not violate SUSY. The terms in the expression for $V_D$ are
proportional to the $SU(2)_L$, $U(1)_Y$ and $U(1)_{N}$ gauge
couplings, {\it i.e.}~$g_2,\,g'$ and $g'_1$, respectively. The values of the
gauge couplings $g_2$ and $g'$ at the EW scale are well known, whereas
the low--energy value of the extra $U(1)_{N}$ coupling $g'_1$ and the
effective $U(1)_{N}$ charges of $H_d$, $H_u$ and $S$ can be calculated
assuming gauge coupling unification \cite{King:2005jy}.  The soft SUSY
breaking terms are collected in $V_{soft}$ and include the soft masses
$m_{H_d}^2,\,m_{H_u}^2,\,m_{S}^2$,\, $m_{\overline{S}}^2$ and $m^2_{\phi}$,
the trilinear couplings $A_{\lambda}$, $A_{\sigma}$ and $A_{\kappa}$,
the bilinear coupling $B$ and a linear coupling
$\xi$. The term $\Delta V$ in Eq.~(\ref{hd9}) contains the loop
corrections to the Higgs effective potential. In SUSY models the most
significant contribution to $\Delta V$ comes from the loops involving
the top-quark and its superpartners, the stops.

At the physical minimum of the scalar potential, Eq.~(\ref{hd9}), the Higgs fields develop VEVs
\be
\ba{c}
<H_d>=\ds\frac{1}{\sqrt{2}}\left(
\begin{array}{c}
v_1\\ 0
\end{array}
\right) , \qquad
<H_u>=\ds\frac{1}{\sqrt{2}}\left(
\begin{array}{c}
0\\ v_2
\end{array}
\right) ,\\[5mm]
<S>=\ds\frac{s_1}{\sqrt{2}} \;,\qquad
<\overline{S}>=\ds\frac{s_2}{\sqrt{2}} \;,\qquad
<\phi>=\ds\frac{\varphi}{\sqrt{2}}\;.
\ea
\label{hd10}
\ee
Using the short-hand notation $\partial
  V / \partial \Phi _{(\Phi = \langle \Phi \rangle)} \equiv \partial
  V / \partial \langle \Phi \rangle$, the minimum conditions for the Higgs potential
of  Eq.~(\ref{hd9}) read,
\be
\ba{rcl}
\dfrac{\partial V}{\partial s_1}&=& m_{S}^2\, s_1 - \dfrac{\lambda A_{\lambda}}{\sqrt{2}} v_1 v_2
- \dfrac{\sigma A_{\sigma}}{\sqrt{2}} \varphi s_2
+ \left(\dfrac{\sigma}{2} s_1 s_2 - \dfrac{\kappa}{2} \varphi^2 - \dfrac{\mu}{\sqrt{2}}\varphi - \Lambda\right)\sigma s_2 \\[2mm]
&+& \dfrac{\sigma^2}{2} \varphi^2 s_1 +
\dfrac{g^{\prime \, 2}_1}{2}\biggl(\tilde{Q}_{H_d} v_1^2+\tilde{Q}_{H_u} v_2^2+\tilde{Q}_S s_1^2 - \tilde{Q}_S s_2^2 \biggr)\tilde{Q}_S s_1 \\[2mm]
&+& \dfrac{\lambda^2}{2} (v_1^2+v_2^2) s_1 + \dfrac{\partial\Delta V}{\partial s_1}=0\,,\\[4mm]
\dfrac{\partial V}{\partial s_2}&=& m_{\overline{S}}^2\, s_2 - \dfrac{\sigma A_{\sigma}}{\sqrt{2}} \varphi s_1
+ \left(\dfrac{\sigma}{2} s_1 s_2 - \dfrac{\kappa}{2} \varphi^2 - \dfrac{\mu}{\sqrt{2}}\varphi - \Lambda\right)\sigma s_1 \\[2mm]
&+& \dfrac{\sigma^2}{2} \varphi^2 s_2 -
\dfrac{g^{\prime \, 2}_1}{2}\biggl(\tilde{Q}_{H_d} v_1^2+\tilde{Q}_{H_u} v_2^2+\tilde{Q}_S s_1^2 - \tilde{Q}_S s_2^2 \biggr)\tilde{Q}_S s_2 \\[2mm]
&+& \dfrac{\lambda \sigma}{2} v_1 v_2 \varphi + \dfrac{\partial\Delta V}{\partial s_2}=0\,,\\[4mm]
\dfrac{\partial V}{\partial \varphi}&=& m_{\varphi}^2\, \varphi - \dfrac{\sigma A_{\sigma}}{\sqrt{2}} s_1 s_2 + B\mu \varphi + \sqrt{2}\xi\Lambda
+ \dfrac{\kappa A_{\kappa}}{\sqrt{2}} \varphi^2 + \dfrac{\sigma^2}{2}(s_1^2+s_2^2)\varphi\\[2mm]
&-& 2 \left(\dfrac{\sigma}{2} s_1 s_2 - \dfrac{\kappa}{2} \varphi^2 - \dfrac{\mu}{\sqrt{2}}\varphi - \Lambda\right)\left(\kappa \varphi +\dfrac{\mu}{\sqrt{2}}\right)
+ \dfrac{\lambda \sigma}{2} v_1 v_2 s_2 + \dfrac{\partial\Delta
  V}{\partial \varphi}=0\,,
%
\ea
\label{hd11}
\ee
\be
\ba{rcl}
\ds\frac{\partial V}{\partial v_1}&=& m_{H_d}^2\, v_1 - \dfrac{\lambda A_{\lambda}}{\sqrt{2}}s_1 v_2 + \dfrac{\lambda \sigma}{2} v_2 s_2 \varphi
+\dfrac{\lambda^2}{2}(v_2^2+s_1^2)v_1+\dfrac{\bar{g}^2}{8}\biggl(v_1^2-v_2^2\biggr)v_1\\[2mm]
&+&\dfrac{g^{\prime \, 2}_1}{2}\biggl(\tilde{Q}_{H_d} v_1^2+\tilde{Q}_{H_u} v_2^2+\tilde{Q}_S s_1^2 - \tilde{Q}_S s_2^2 \biggr)\tilde{Q}_{H_d} v_1
+\dfrac{\partial\Delta V}{\partial v_1}=0\,,\\[4mm]
\ds\frac{\partial V}{\partial v_2}&=& m_{H_u}^2 v_2-\dfrac{\lambda A_{\lambda}}{\sqrt{2}}s_1 v_1 + \dfrac{\lambda \sigma}{2} v_1 s_2 \varphi
+\dfrac{\lambda^2}{2}(v_1^2+s_1^2)v_2+\dfrac{\bar{g}^2}{8}\biggl(v_2^2-v_1^2\biggr)v_2\\[2mm]
&&+\ds\frac{g^{\prime \,2}_1}{2}\biggl(\tilde{Q}_{H_d}v_1^2+\tilde{Q}_{H_u}v_2^2+\tilde{Q}_S s_1^2 - \tilde{Q}_S s_2^2 \biggr)\tilde{Q}_{H_u} v_2
+ \ds\frac{\partial\Delta V}{\partial v_2}=0\,,
\ea
\nonumber
\ee
where $\bar{g}=\sqrt{g_2^2+g'^2}$. Instead of specifying $v_1$,
$v_2$, $s_1$ and $s_2$, it is more convenient to use
\begin{equation}
\tan\beta=v_2/v_1 \qquad \mbox{and} \qquad \tan\theta=s_2/s_1 \;.
\end{equation}
The VEV $v$ is given by the electroweak scale,
  $v=\sqrt{v_1^2+v_2^2} \approx 246\,\mbox{GeV}$, and
  $s=\sqrt{s_1^2+s_2^2}$ sets the $Z^\prime$ mass, as discussed below.

Initially the Higgs sector involves fourteen degrees of freedom. However four of them are massless Goldstone modes.
They are swallowed by the $W^{\pm}$, $Z$ and $Z'$ gauge bosons. The charged $W^{\pm}$ bosons gain masses
via the interaction with the neutral components of the Higgs doublets $H_u$ and $H_d$ just in the same way as in the
MSSM, resulting in $M_W=\dfrac{g_2}{2}v$. On the other hand the mechanism of the neutral gauge boson mass
generation differs substantially. Let the $Z'$ and $Z$ states be the gauge bosons associated with the group $U(1)_{N}$ and
with the SM-like $Z$ boson, respectively. Then the $Z-Z'$ mass-squared matrix is given by
\be
M^2_{ZZ'}=\left(
\ba{cc}
\dfrac{\bar{g}^2}{4} v^2  & \Delta^2 \\[2mm]
\Delta^2  & g^{\prime \,2}_1v^2\biggl(\tilde{Q}_{H_d}^2\cos^2\beta+\tilde{Q}_{H_u}^2\sin^2\beta\biggr) + g^{\prime \,2}_1 \tilde{Q}^2_S s^2
\ea
\right)\,,
\label{hd12}
\ee
where
$$
\Delta^2=\ds\frac{\bar{g}g'_1}{2}v^2\biggl(\tilde{Q}_{H_d}\cos^2\beta-\tilde{Q}_{H_u}\sin^2\beta\biggr)\,.
$$
The fields $S$ and $\overline{S}$ must acquire large VEVs, {\it
  i.e.}~$s_1\simeq s_2 \gg 1\,\mbox{TeV}$, to ensure
that the extra $U(1)_{N}$ gauge boson is sufficiently heavy. Then the mass of the lightest neutral gauge boson $Z_1$ is very close
to $M_Z=\bar{g}v/2$, whereas the mass of $Z'$ is set by
$M_{Z'}\approx  g'_1\tilde{Q}_S\, s$.

\subsection{The Higgs Boson Spectrum}

For the analysis of the Higgs boson spectrum we use Eq.~(\ref{hd11})
for the extrema to express the soft masses
$m_{H_d}^2,\,m_{H_u}^2,\,m_{S}^2$,\, $m_{\overline{S}}^2$ and $m^2_{\phi}$ in
terms of $s,\,v,\,\varphi,\, \beta,\,
\theta$ and other parameters. Because of the conversation of the
electric charge, the charged components of the Higgs doublets are not
mixed with the neutral Higgs fields. They form a separate sector, the
spectrum of which is described by a $2\times 2$ mass matrix. The determinant of
this matrix is zero and results in the appearance of two Goldstone
states, {\it i.e.}
\be
G^-=H_d^{-}\cos\beta-H_u^{+*}\sin\beta
\label{hd13}
\ee
and its charge conjugate (that are absorbed into the longitudinal
degrees of freedom of the $W^{\pm}$ gauge bosons) and of two charged
Higgs states,
\be
H^{+}=H_d^{-*}\sin\beta+H_u^{+}\cos\beta
\label{hd14} \,
\ee
with mass
\be
m^2_{H^{\pm}}=\dfrac{\sqrt{2}\lambda s}{\sin 2\beta}\left(A_{\lambda} \cos\theta - \dfrac{\sigma \varphi}{\sqrt{2}}\sin\theta\right)
-\frac{\lambda^2}{2}v^2+\frac{g_2^2}{4}v^2+\Delta_{\pm}\,,
\label{hd15}
\ee
where $\Delta_{\pm}$ denotes the loop corrections to $m^2_{H^{\pm}}$.

The imaginary parts of the neutral components of the Higgs doublets and the imaginary parts of the two SM singlet fields
$S$ and $\overline{S}$ compose two neutral Goldstone states
\be
\ba{l}
G=\sqrt{2}(\mbox{Im}\,H_d^0\cos\beta - \mbox{Im}\, H_u^0\sin\beta)\,,\\[1mm]
G'=\sqrt{2}(\mbox{Im}\,S \cos\theta - \mbox{Im}\,\overline{S}\sin\theta)\cos\gamma - \sqrt{2}(\mbox{Im}\,H_u^0\cos\beta + \mbox{Im}\,H_d^0\sin\beta)\sin\gamma\,,
\ea
\label{hd16}
\ee
which are swallowed by the $Z$ and $Z'$ bosons, as well as three physical states.
In Eq.~(\ref{hd16}) we have introduced
\begin{equation}
\tan\gamma=\dfrac{v}{2s}\sin 2\beta \;.
\end{equation}
In the field basis $(P_1,\,P_2,\,P_3)$, where
\be
\ba{l}
P_1=\sqrt{2}(\mbox{Im}\,H_u^0\cos\beta + \mbox{Im}\,H_d^0\sin\beta)\cos\gamma + \sqrt{2}(\mbox{Im}\,S \cos\theta - \mbox{Im}\,\overline{S}\sin\theta)\sin\gamma\,, \\[1mm]
P_2=\sqrt{2}\left(\mbox{Im}\,S\sin\theta + \mbox{Im}\,\overline{S}\cos\theta\right)\,,\\[1mm]
P_3=\sqrt{2} \mbox{Im}\,\phi\,,
\ea
\label{hd17}
\ee
the mass matrix of the CP-odd Higgs sector takes the form
\be
\tilde{M}^2= (\tilde{M}^2_{ij}) \;,  \quad  i,j=1,2,3 \;,
\label{hd18}
\ee
with
\be
\ba{rcl}
\tilde{M}_{11}^2&=&\dfrac{\sqrt{2}\lambda s}{\sin 2\beta \cos^2 \gamma}\left(A_{\lambda} \cos\theta - \dfrac{\sigma \varphi}{\sqrt{2}}\sin\theta\right)+\tilde{\Delta}_{11}\,,\\[3mm]
\tilde{M}_{12}^2&=&\tilde{M}_{21}^2=\dfrac{\lambda
  v}{\sqrt{2}\cos\gamma}\left(A_{\lambda} \sin\theta + \dfrac{\sigma
    \varphi}{\sqrt{2}}\cos\theta\right)+\tilde{\Delta}_{12}\,,\\[3mm]
\tilde{M}_{13}^2&=&\tilde{M}_{31}^2=\dfrac{\lambda\sigma v s}{2\cos
  \gamma}\sin\theta+\tilde{\Delta}_{13}\,,
\\[3mm]
\tilde{M}_{22}^2&=&\dfrac{2\sigma\varphi}{\sin 2\theta}\left(\dfrac{A_{\sigma}}{\sqrt{2}}+\dfrac{\kappa}{2}\varphi+\dfrac{\mu}{\sqrt{2}}+\dfrac{\Lambda}{\varphi}\right)\\[3mm]
&+&\dfrac{\lambda v^2 \sin 2\beta}{\sqrt{2} s \sin 2\theta}\left(A_{\lambda}\sin^3\theta-\dfrac{\sigma\varphi}{\sqrt{2}}\cos^3\theta\right)+\tilde{\Delta}_{22}\,,\\[3mm]
\tilde{M}_{23}^2&=&\tilde{M}_{32}^2=\sigma s \left(\dfrac{A_{\sigma}}{\sqrt{2}}-\kappa \varphi -\dfrac{\mu}{\sqrt{2}} \right) -\dfrac{\lambda\sigma}{4}v^2 \sin 2\beta \cos\theta
+\tilde{\Delta}_{23}\,,
\ea
\label{hd19}
\ee
\be
\ba{rcl}
\tilde{M}_{33}^2&=&\dfrac{\sigma s^2}{2\sqrt{2}\varphi} A_{\sigma} \sin 2\theta - 2 B\mu -3\dfrac{\kappa A_{\kappa}}{\sqrt{2}}\varphi
- \sqrt{2}(\xi+\mu)\dfrac{\Lambda}{\varphi} + \sigma\kappa s^2 \sin 2\theta - \dfrac{\kappa\mu}{\sqrt{2}}\varphi \\[3mm]
&-&4\kappa\Lambda + \dfrac{\sigma\mu s^2}{2\sqrt{2}\varphi} \sin 2\theta - \dfrac{\lambda\sigma s}{4\varphi} v^2 \sin\theta \sin 2\beta+\tilde{\Delta}_{33}\,.
\ea
\nonumber
\ee
In Eqs.~(\ref{hd19}) the $\tilde{\Delta}_{ij}$ ($i,j=1,2,3$) denote loop
corrections. Since in the models under consideration $s$ must be much
larger than $v$, it follows that $\gamma$ goes to zero. Moreover,
since in phenomenologically acceptable SUSY models the supersymmetry
breaking scale also tends to be considerably larger than $v$, the
mixing between $P_1$ and the other two pseudoscalar states $P_2$ and
$P_3$ is somewhat suppressed. So one CP-odd mass eigenstate is predominantly
$P_1$. The other two CP-odd mass eigenstates are mainly made up of linear
superpositions of the imaginary parts of the SM singlet fields $S$,
$\overline{S}$ and $\phi$, {\it i.e.}~of $P_2$ and $P_3$. In other words, as
the off-diagonal entries $\tilde{M}_{12}^2,\, \tilde{M}_{13}^2\ll
\tilde{M}_{11}^2$, the mass matrix, Eqs.~(\ref{hd18}) and (\ref{hd19}), can be
diagonalised analytically.  In
particular, the mass of the CP-odd Higgs eigenstate, that is
predominantly $P_1$, is set by $\tilde{M}_{11}^2$. As a consequence
this CP-odd state and the charged physical Higgs states are expected
to be approximately degenerate.

The mass matrix, Eqs.~(\ref{hd18}) and (\ref{hd19}), is diagonalised by
means of a unitary transformation $U$ that relates the components of the CP-odd
Higgs basis Eq.~(\ref{hd17}) to the corresponding Higgs mass
eigenstates $A_i$ ($i=1,2,3$),
\be
\left(
\begin{array}{c}
P_1\\ P_2\\ P_3
\end{array}
\right)=
U
\left(
\begin{array}{c}
A_1 \\ A_2\\ A_3
\end{array}
\right)\,.
\label{hd20}
\ee
The pseudoscalar mass eigenstates are labeled according to increasing absolute
value of mass, where $A_1$ is the lightest CP-odd Higgs state and
$A_3$ the heaviest.  At tree-level and neglecting all terms
proportional to $\lambda v$ one obtains
\be
\ba{lcl}
m^2_{A_3} & \simeq & \mbox{max} \biggl\{\dfrac{\sqrt{2}\sigma A_{\sigma} \varphi}{\sin 2\theta \cos^2 \delta}\,,\quad
\dfrac{\sqrt{2}\lambda s}{\sin 2\beta}\left(A_{\lambda} \cos\theta - \dfrac{\sigma \varphi}{\sqrt{2}}\sin\theta\right) \biggr\}\,,\\[3mm]
m^2_{A_2} & \simeq & \mbox{min} \biggl\{\dfrac{\sqrt{2}\sigma A_{\sigma} \varphi}{\sin 2\theta \cos^2 \delta}\,,\quad
\dfrac{\sqrt{2}\lambda s}{\sin 2\beta}\left(A_{\lambda} \cos\theta - \dfrac{\sigma \varphi}{\sqrt{2}}\sin\theta\right)\biggr\}\,,\\[3mm]
m^2_{A_1} & \simeq & \cos^2 \delta \left[- 2 B\mu -3\dfrac{\kappa A_{\kappa}}{\sqrt{2}}\varphi
- \sqrt{2}\xi \dfrac{\Lambda}{\varphi} + \dfrac{9}{4} \sigma\kappa s^2 \sin 2\theta \right.\\[3mm]
&&\left. + \sqrt{2} \dfrac{\sigma\mu s^2}{\varphi} \sin 2\theta + \dfrac{\sigma s^2 \Lambda}{2 \varphi^2} \sin 2\theta\right]\,,
\ea
\label{hd21}
\ee
where we have defined
\begin{equation}
\tan\delta \simeq \dfrac{s}{2\varphi}\sin 2\theta \;.
\end{equation}
In this case the lightest CP-odd mass eigenstate is a linear combination of
$P_2$ and $P_3$,
 \be
A_1 \simeq  - P_2 \sin\delta + P_3 \cos\delta\,.
\label{hd22}
\ee
In the limit where the previously discussed global $U(1)$ symmetry
violating couplings $\kappa$, $\mu$ and $\Lambda$ vanish, the mass of
the lightest CP--odd Higgs boson goes to zero.

The CP-even Higgs sector involves the
$\mbox{Re}\,H_d^0$, $\mbox{Re}\,H_u^0$, $\mbox{Re}\,S$, $\mbox{Re}\,\overline{S}$ and $\mbox{Re}\,\phi$.
In the field space basis $(S_1,\,S_2,\,S_3,\,S_4,\,S_5)$, where
\be
\ba{lcl}
\mbox{Re}\,S & = & (S_1\,\cos\theta + S_2\sin\theta + s_1)/\sqrt{2}\,, \\[2mm]
\mbox{Re}\,\overline{S} & = & (-S_1\,\sin\theta + S_2\cos\theta + s_2)/\sqrt{2}\,, \\[2mm]
\mbox{Re}\,\phi & = & (S_3 + \varphi)/\sqrt{2}\,,\\[2mm]
\mbox{Re}\,H_d^0 & = &(S_5 \cos\beta - S_4 \sin\beta+v_1)/\sqrt{2}\,, \\[2mm]
\mbox{Re}\,H_u^0 & = &(S_5 \sin\beta + S_4 \cos\beta+v_2)/\sqrt{2}\,, \\[2mm]
\ea
\label{hd23}
\ee
the mass matrix of the CP-even Higgs sector takes the form
\be
M^2= (M^2_{ij}) \;, \quad  i,j=1,...,5 \;,
\label{hd24}
\ee
where
\be
\ba{rcl}
M_{11}^2& = & g^{\prime \,2}_1 \tilde{Q}_S^2 s^2 - \dfrac{\sigma^2 s^2}{2} \sin^2 2\theta + \sqrt{2} \sigma A_{\sigma} \varphi \sin 2\theta\\[2mm]
& + &\biggl(\kappa \sigma \varphi^2 + \sqrt{2} \sigma \mu \varphi + 2\sigma \Lambda\biggr) \sin 2\theta
+ \dfrac{\lambda A_{\lambda}}{2\sqrt{2} s} v^2 \cos\theta \sin 2\beta\\[2mm]
& - & \dfrac{\lambda \sigma \varphi}{4 s} v^2 \sin\theta \sin 2\beta + \Delta_{11}\,,\\[2mm]
M_{12}^2& = &M_{21}^2=\dfrac{\sigma^2 s^2}{4} \sin 4\theta - \sqrt{2} \sigma A_{\sigma} \varphi \cos 2\theta \\[2mm]
& - & \biggl(\kappa \sigma \varphi^2 + \sqrt{2} \sigma \mu \varphi + 2\sigma \Lambda\biggr) \cos 2\theta
+ \dfrac{\lambda A_{\lambda}}{2\sqrt{2} s} v^2 \sin \theta \sin 2\beta \\[2mm]
& + & \dfrac{\lambda \sigma \varphi}{4 s} v^2 \cos\theta \sin 2\beta + \Delta_{12}\,,\\[2mm]
M_{13}^2& = & M_{31}^2= \sigma^2 \varphi s \cos 2\theta -
\dfrac{\lambda\sigma}{4} v^2\sin\theta\sin 2\beta +
\Delta_{13}\,,\\[2mm]
M_{14}^2& = & M_{41}^2=\dfrac{g^{\prime \,2}_1}{2}\tilde{Q}_S(\tilde{Q}_{H_u}-\tilde{Q}_{H_d}) s v \sin 2\beta
- \dfrac{\lambda A_{\lambda}}{\sqrt{2}} v \cos\theta \cos 2\beta\\[2mm]
& - & \dfrac{\lambda\sigma}{2}\varphi v \sin\theta \cos 2\beta +
\Delta_{14}\,,\\[2mm]
M_{15}^2& = & M_{51}^2=g^{\prime \,2}_1 \tilde{Q}_S (\tilde{Q}_{H_d}\cos^2\beta+ \tilde{Q}_{H_u}\sin^2\beta) s v
-\dfrac{\lambda A_{\lambda}}{\sqrt{2}} v \cos\theta \sin 2\beta\,,\\[2mm]
&+ & \lambda^2 v s \cos^2 \theta - \dfrac{\lambda\sigma}{2} \varphi v \sin\theta \sin 2\beta + \Delta_{15}\,,\\[2mm]
M_{22}^2& = & \dfrac{\sigma^2 s^2}{2} \sin^2 2\theta + \dfrac{\sqrt{2} \sigma A_{\sigma} \varphi}{\sin 2\theta} \cos^2 2\theta
+ \biggl(\kappa \sigma \varphi^2 + \sqrt{2} \sigma \mu \varphi + 2\sigma \Lambda\biggr) \dfrac{\cos^2 2\theta}{\sin 2\theta}\\[2mm]
&+&\dfrac{\lambda A_{\lambda} v^2}{2\sqrt{2} s \cos\theta} \sin^2 \theta \sin 2\beta
- \dfrac{\lambda \sigma \varphi v^2}{4 s \sin\theta} \cos^2 \theta \sin 2\beta + \Delta_{22}\,,\\[2mm]
M_{23}^2& = & M_{32}^2=-\dfrac{\sigma A_{\sigma}}{\sqrt{2}}s+\sigma^2\varphi s \sin 2\theta
- \sigma s \left(\kappa\varphi+\dfrac{\mu}{\sqrt{2}}\right)\\[2mm]
&+ &\dfrac{\lambda\sigma}{4}v^2 \cos\theta \sin 2\beta +
\Delta_{23}\,,
\nonumber
\ea
\ee
\be
\ba{rcl}
M_{24}^2& = & M_{42}^2=\left(-\dfrac{\lambda A_{\lambda}}{\sqrt{2}} v \sin\theta +
\dfrac{\lambda\sigma}{2} \varphi v \cos\theta \right)\cos 2\beta + \Delta_{24}\,,\\[2mm]
M_{25}^2& = & M_{52}^2=\dfrac{\lambda^2}{2} s v \sin 2\theta + \left(-\dfrac{\lambda A_{\lambda}}{\sqrt{2}} v \sin\theta +
\dfrac{\lambda\sigma}{2} \varphi v \cos\theta \right)\sin 2\beta +
\Delta_{25}\,,\\[2mm]
M_{33}^2& = &\dfrac{\sigma A_{\sigma} s^2}{2\sqrt{2}\varphi} \sin 2\theta - \sqrt{2}\xi\dfrac{\Lambda}{\varphi} + \dfrac{\kappa A_{\kappa}}{\sqrt{2}} \varphi
+\mu\left(\dfrac{\sigma s^2}{2\sqrt{2}\varphi}\sin 2\theta+3\dfrac{\kappa\varphi}{\sqrt{2}}-\dfrac{\sqrt{2}\Lambda}{\varphi}\right) \\[2mm]
&+ &2 \kappa^2 \varphi^2 -\dfrac{\lambda\sigma s}{4\varphi} v^2
\sin\theta \sin 2\beta + \Delta_{33}\,,
\\[2mm]
M_{34}^2& = & M_{43}^2=\dfrac{\lambda\sigma}{2} s v \sin\theta \cos 2\beta + \Delta_{34}\,,\\[2mm]
M_{35}^2& = & M_{53}^2=\dfrac{\lambda\sigma}{2} s v \sin\theta \sin 2\beta + \Delta_{35}\,,\\[2mm]
M_{44}^2&=&\dfrac{\sqrt{2}\lambda s}{\sin 2\beta}\left(A_{\lambda} \cos\theta - \dfrac{\sigma \varphi}{\sqrt{2}}\sin\theta\right)+
\left(\dfrac{\bar{g}^2}{4}-\dfrac{\lambda^2}{2}\right)v^2 \sin^2 2\beta\\[2mm]
&+&\ds\frac{g^{\prime \,2}_1}{4}(\tilde{Q}_{H_u}-\tilde{Q}_{H_d})^2 v^2 \sin^2 2\beta+\Delta_{44}\,,\\[2mm]
M_{45}^2&=&M_{54}^2=\left(\dfrac{\lambda^2}{4}-\dfrac{\bar{g}^2}{8}\right)v^2 \sin 4\beta
+\dfrac{g^{\prime \,2}_1}{2}v^2(\tilde{Q}_{H_u}-\tilde{Q}_{H_d})\times\\[2mm]
&\times &(\tilde{Q}_{H_d}\cos^2\beta+\tilde{Q}_{H_u}\sin^2\beta)\sin 2\beta+\Delta_{45}\, ,\\[2mm]
M_{55}^2&=&\dfrac{\lambda^2}{2}v^2\sin^22\beta+\dfrac{\bar{g}^2}{4}v^2\cos^22\beta+g^{\prime \,2}_1 v^2(\tilde{Q}_{H_d}\cos^2\beta+
\tilde{Q}_{H_u}\sin^2\beta)^2+\Delta_{55}\,.
\ea
\label{hd25}
\ee
In Eq.~(\ref{hd25}) the $\Delta_{ij}$ denote the loop corrections. The
components of the CP-even Higgs basis, Eq.~(\ref{hd23}), are related
to the CP-even Higgs mass eigenstates $h_i$ ($i=1,...,5$) by
virtue of a unitary transformation,
\be
\left(
\begin{array}{c}
S_1\\S_2\\ S_3\\ S_4\\ S_5
\end{array}
\right)=
\tilde{U}
\left(
\begin{array}{c}
h_1 \\ h_2\\ h_3 \\ h_4\\ h_5
\end{array}
\right)\,,
\label{hd26}
\ee
where again the CP-even Higgs eigenstates are labeled according to
increasing absolute value of mass, with $h_1$ being the lightest
CP-even Higgs state and $h_5$ the heaviest.

If all SUSY breaking parameters as well as $\lambda s \sim \sigma s
\sim \sigma \varphi \sim M_{S}$ are considerably larger than the EW
scale, all masses of the CP-even Higgs states except for the lightest
Higgs boson mass are determined by the SUSY breaking scale $M_S$.
Because the minimal eigenvalue of the mass matrix, Eqs.~(\ref{hd24})--(\ref{hd25}),
is always less than its smallest diagonal element the lightest Higgs
state in the CP-even sector (approximately $S_5$) remains always light
irrespective of the SUSY breaking scale, {\it i.e.}~$m^2_{h_1}\lesssim
M_{55}^2$ like in the MSSM and NMSSM.  In the interactions with other
SM particles this state manifests itself as a SM-like Higgs boson if $M_S \gg M_Z$.

In the limit where $\lambda \sim \sigma \to 0 $, the off-diagonal
tree-level entries $M_{24}^2$, $M_{25}^2$, $M_{34}^2$ and $M_{35}^2$
of the mass matrix, Eqs.~(\ref{hd24})--(\ref{hd25}), become negligibly
small.  At the same time, according to Eq.~(\ref{hd3}), the diagonal
entry $M_{11}^2$ that is set by the mass of the $Z'$ boson tends to be
substantially larger than $M_S^2$, {\it i.e.}~$M_{11}^2\simeq M_{Z'}^2 \sim
M_S^2/\sigma^2$, whereas $\cos 2\theta$ almost vanishes in this
case. Indeed, combining the first and the second equations for the
extrema Eq.~(\ref{hd11}) one obtains the following tree-level
expression for $\cos 2\theta$
\be \cos 2\theta
\simeq
\dfrac{m_{\overline{S}}^2-m_{S}^2}{m_{\overline{S}}^2+m_S^2+\sigma^2\varphi^2+g^{\prime \,2}_1
  \tilde{Q}_S^2 s^2} \sim \dfrac{M_S^2}{M_{Z'}^2}\sim \sigma^2\,,
\label{hd27}
\ee
which becomes vanishingly small when $M_{Z'}\gg M_S$ and/or when
$\sigma\sim \lambda\to 0$. In this case the  hierarchical structure
of the mass matrix, Eqs.~(\ref{hd24})--(\ref{hd25}), implies that the
mass of the $Z'$ boson  and the mass of the heaviest CP-even Higgs particle
associated with $S_1$ are almost degenerate. Thus the heaviest CP-even
Higgs state can be integrated out. The mass of another CP-even state
that is predominantly $S_4$, the mass of the CP-odd state that
corresponds to $P_1$, and the masses of the charged Higgs states
are also almost degenerate in this limit. Assuming that the Higgs state that
 is mainly $S_4$ is the second heaviest CP-even Higgs
state, neglecting all terms that are proportional to the global
$U(1)$-violating couplings ($\kappa$, $\mu$  and $\Lambda$) and
setting $\cos 2\theta=0$, one obtains the following approximate
analytic expressions for the tree-level masses of the three lightest
CP-even Higgs bosons,
\be
\ba{lcl}
m_{h_{3,2}}^2 & \simeq & \dfrac{\sigma^2 s^2}{4}\left[1+\dfrac{A_{\sigma}}{\sqrt{2}\sigma \varphi} \pm
\biggl|1 - \dfrac{A_{\sigma}}{\sqrt{2}\sigma \varphi}\biggr|\sqrt{1+16\,\dfrac{\varphi^2}{s^2}}\right]\,,\\[2mm]
m_{h_{1}}^2 & \simeq & \dfrac{\bar{g}^2}{4}v^2\cos^22\beta\simeq M_{Z}^2 \cos^2 2\beta\,.
\ea
\label{hd28}
\ee
Note that in the scenario under consideration the
tree-level expression for the SM-like Higgs mass $m_{h_{1}}^2$ is
essentially the same as in the MSSM.

\section{Non-Standard Higgs Decays}

We now focus on that region of the parameter space that corresponds to the
approximate global $U(1)$ symmetry mentioned above, where we have a
light pseudoscalar. Since our primary concern in our numerical
investigation is to study non-standard Higgs decays and we do not
assume any breaking pattern among the soft masses, most of the
sfermion masses do not play a significant role. We therefore choose the SUSY
breaking parameters that control the masses of the sfermions to be
well above the TeV scale thus comfortably evading limits set by the
LHC and decoupling them from the spectrum. We do, however, adjust the
stop mass parameters to get a Higgs mass of $125-126$ GeV.  The
gaugino masses are also chosen to be heavy enough to give a Higgsino dark
matter candidate and to evade the LHC limit on the gluino mass.

 Additionally we assume that the SM singlet superfields
 $\overline{S}$, $S$ and $\phi$ acquire very large VEVs inducing
 multi-TeV masses of the $Z'$ boson. Our analysis of the Higgs
 sector in the previous section indicates that the Higgs spectrum in
 general has a very hierarchical structure when all SUSY breaking
 parameters are sufficiently large, i.e.~above about $1\,\mbox{TeV}$. In
 this limit only the SM-like Higgs boson and the lightest CP-odd
 Higgs state associated with the spontaneously broken approximate
 global $U(1)$ symmetry can be considerably lighter than
 $1\,\mbox{TeV}$.

The presence of a light pseudoscalar Higgs state in the particle
spectrum can result in non-standard decays of the lightest CP-even
Higgs boson in the $U(1)$ extensions of the MSSM under consideration. Expanding
the Higgs potential, Eq.~(\ref{hd9}), about its physical minimum one
obtains the trilinear coupling that describes the interaction of the
lightest CP-even Higgs scalar with the lightest pseudoscalar Higgs
states. At  tree-level the corresponding part of the Lagrangian can be written as
\begin{equation}
\begin{array}{l}
\mathcal{L}_{h_1 A_1 A_1}=-G_{h_1 A_1 A_1} h_1 A_1 A_1\,,
\end{array}
\label{hd29}
\end{equation}
with the trilinear Higgs couplings $G_{h_1 A_1 A_1}$ which is given by
the rather lengthy expression
\begin{equation}
\begin{array}{l}
G_{h_1 A_1 A_1}=\tilde{U}_{51}\left\{U_{11}^2 \left[\dfrac{\lambda^2}{4}v\cos^2\gamma (1+\cos^2 2\beta) + \dfrac{\lambda^2}{2} v \sin^2 \gamma \cos^2 \theta
-\dfrac{\bar{g}^2}{8} v \cos^2 \gamma \cos^2 2\beta \right.\right.
\\[1mm]
\left.\left.+\dfrac{1}{2}\left(\dfrac{\lambda A_{\lambda}}{\sqrt{2}} \cos\theta -
\dfrac{\lambda \sigma}{2} \varphi \sin\theta\right)\sin 2\gamma+
\dfrac{g_1^{\prime \,2}}{2}v\left(\tilde{Q}_{H_d}\cos^2\beta+\tilde{Q}_{H_u}\sin^2\beta\right)\times\right.\right.
\\[1mm]
\left.\left.\times\left(\tilde{Q}_{H_d}\sin^2\beta \cos^2\gamma +
\tilde{Q}_{H_u}\cos^2\beta \cos^2\gamma +\tilde{Q}_{S}\sin^2\gamma
\cos 2\theta\right)\right]\right.
\\[1mm]
\left.+ U_{11} U_{21} \left[\dfrac{\lambda^2}{2} v \sin 2\theta \sin \gamma + g_1^{\prime \,2} \tilde{Q}_S v \left(\tilde{Q}_{H_d}\cos^2\beta
+\tilde{Q}_{H_u}\sin^2\beta\right) \sin\gamma \sin 2\theta
\right.\right.
\\[1mm]
\left.\left. + \left(\dfrac{\lambda A_{\lambda}}{\sqrt{2}}\sin\theta + \dfrac{\lambda \sigma}{2} \varphi \cos\theta\right) \cos\gamma\right]
+ \dfrac{\lambda\sigma}{2} \sin\theta\, U_{11} U_{31} (s\cos\gamma + v
\sin 2\beta \sin\gamma)\right.
\\[1mm]
\left. + U_{21}^2\left[\dfrac{\lambda^2}{2}v\sin^2\theta - \dfrac{g_1^{\prime \,2}}{2} \tilde{Q}_S v \cos 2\theta
\left(\tilde{Q}_{H_d}\cos^2\beta+\tilde{Q}_{H_u}\sin^2\beta\right)\right]
\right.
\\[1mm]
\left.- \dfrac{\lambda\sigma}{2} v \sin 2\beta \cos\theta\, U_{21} U_{31} \right\}
+\tilde{U}_{41}\left\{U_{11}^2
  \left[\left(-\dfrac{\lambda^2}{8}+\dfrac{\bar{g}^2}{16}\right)v\cos^2\gamma
    \sin 4\beta\right.\right.
\\[1mm]
\left.\left.+\dfrac{g_1^{\prime \,2}}{4}v\sin 2\beta (\tilde{Q}_{H_u}-\tilde{Q}_{H_d})\left(\tilde{Q}_{H_d}\sin^2\beta \cos^2\gamma +
\tilde{Q}_{H_u}\cos^2\beta \cos^2\gamma \right.\right.\right.
\\[1mm]
\left.\left.\left.+\tilde{Q}_{S}\sin^2\gamma \cos 2\theta\right)\right]+\dfrac{g_1^{\prime \,2}}{2}\tilde{Q}_{S} (\tilde{Q}_{H_u}-\tilde{Q}_{H_d})
v\sin 2\beta \sin\gamma \sin 2\theta\, U_{11} U_{21} \right.
\\[1mm]
\left.+ \dfrac{\lambda\sigma}{2}v\cos 2\beta \sin\gamma \sin\theta\, U_{11} U_{31} - \dfrac{g_1^{\prime \,2}}{4}\tilde{Q}_{S} (\tilde{Q}_{H_u}-\tilde{Q}_{H_d})
v\sin 2\beta \cos 2\theta\, U_{21}^2 \right.
\\[1mm]
\left.- \dfrac{\lambda\sigma}{2}v\cos 2\beta \cos\theta\, U_{21} U_{31}\right\}
+\tilde{U}_{31}\left\{U_{11}^2
  \left[-\dfrac{\lambda\sigma}{4}s\sin\theta \sin 2\beta \cos^2\gamma
    + \dfrac{\sigma^2}{2}\varphi\sin^2 \gamma  \right.\right.
\\[1mm]
\left.\left.-\dfrac{\lambda\sigma}{4} v \sin 2\gamma \sin\theta
- \dfrac{\sigma}{2}\sin 2\theta \sin^2\gamma\left(\dfrac{A_{\sigma}}{\sqrt{2}}+\kappa\varphi
+\dfrac{\mu}{\sqrt{2}}\right)\right] \right.
\\[1mm]
\left.+ U_{11} U_{21} \left[\dfrac{\lambda\sigma}{2}v\cos\theta\cos\gamma
+\sigma\left(\dfrac{A_{\sigma}}{\sqrt{2}}+\kappa\varphi+\dfrac{\mu}{\sqrt{2}}\right)\sin\gamma\cos
2\theta\right] \right.
\nonumber
\end{array}
\end{equation}
\begin{equation}
\begin{array}{l}
\left.+U_{21}^2 \left[\dfrac{\sigma^2}{2}\varphi+\dfrac{\sigma}{2}\left(\dfrac{A_{\sigma}}{\sqrt{2}}+\kappa\varphi
+\dfrac{\mu}{\sqrt{2}}\right)\sin 2\theta\right]-\sigma\kappa s\,
U_{21} U_{31}\right.
\\[1mm]
\left.+\kappa U_{31}^2 \left(\kappa\varphi
+\dfrac{\mu}{\sqrt{2}} - \dfrac{A_{\kappa}}{\sqrt{2}}\right)
\right\}+\tilde{U}_{21}\left\{U_{11}^2
  \left[-\dfrac{\lambda\sigma}{4}\varphi \sin 2\beta \cos^2\gamma
    \cos\theta \right.\right.
\\[1mm]
\left.\left.+ \dfrac{\lambda^2}{4}s\cos^2 \gamma \sin 2\theta + \dfrac{\lambda A_{\lambda}}{2\sqrt{2}}\sin 2\beta \cos^2 \gamma \sin\theta
+\dfrac{\sigma^2}{4}s \sin^2 \gamma \sin 2\theta\right]\right.
\\[1mm]
\left. + U_{11} U_{31}\left[\dfrac{\lambda\sigma}{2}v\cos\gamma\cos\theta+\sigma\left(\dfrac{A_{\sigma}}{\sqrt{2}}-\kappa\varphi
-\dfrac{\mu}{\sqrt{2}}\right) \sin\gamma \cos 2\theta \right]\right.
\\[1mm]
\left. + \dfrac{\sigma^2}{4}s\sin 2\theta\, U_{21}^2+\sigma\left(\dfrac{A_{\sigma}}{\sqrt{2}}-\kappa\varphi
-\dfrac{\mu}{\sqrt{2}}\right)\sin 2\theta\, U_{21}
U_{31}+\dfrac{\sigma}{2}\left(\sigma s \sin 2\theta + \kappa s\right)
U_{31}^2 \right\}
\\[1mm]
+\tilde{U}_{11}\left\{U_{11}^2 \left[\dfrac{\lambda\sigma}{4}\varphi \sin 2\beta \cos^2\gamma \sin\theta
+ \dfrac{\lambda^2}{2}s\cos^2 \gamma \cos^2 \theta + \dfrac{\lambda
  A_{\lambda}}{2\sqrt{2}}\sin 2\beta \cos^2 \gamma
\cos\theta\right.\right.
\\[1mm]
\left.\left.+\dfrac{g_1^{\prime \,2}}{2} \tilde{Q}_{S} s \left(\tilde{Q}_{H_d}\sin^2\beta \cos^2\gamma +
\tilde{Q}_{H_u}\cos^2\beta \cos^2\gamma + \tilde{Q}_{S}\sin^2\gamma
\cos 2\theta\right)\right]\right.
\\[1mm]
\left.+\left[-\dfrac{\lambda\sigma}{2}v\cos\gamma\sin\theta+\sigma\left(\kappa\varphi+\dfrac{\mu}{\sqrt{2}}
-\dfrac{A_{\sigma}}{\sqrt{2}}\right)\sin\gamma \sin 2\theta \right]
U_{11} U_{31} \right.
\\[1mm]
\left. + \left( g_1^{\prime \,2} \tilde{Q}_{S}^2 - \dfrac{\sigma^2}{2} \right) s \sin\gamma \sin 2\theta\, U_{11} U_{21} +
\left[ \dfrac{\sigma^2}{2} - \dfrac{g_1^{\prime \,2}}{2}
  \tilde{Q}_{S}^2\right] s \cos 2\theta\, U_{21}^2\right.
\\[1mm]
\left.+ \sigma\left(\dfrac{A_{\sigma}}{\sqrt{2}}-\kappa\varphi-\dfrac{\mu}{\sqrt{2}}\right)\cos 2\theta\, U_{21} U_{31}
+\dfrac{\sigma^2}{2}s\cos 2\theta\, U_{31}^2\right\}\,.
\end{array}
\label{hd30}
\end{equation}
If $m_{A_1} \lesssim 60$ GeV then the CP-even Higgs boson with
  mass around 125 GeV can decay into a pair of two lightest
  pseudoscalar Higgs bosons $A_1$, through the interaction given in
  Eq.~(\ref{hd29}).  The corresponding partial decay width is given by
\begin{equation}
\Gamma(h_1\to A_1 A_1)=\dfrac{G_{h_1 A_1 A_1}^2}{8\pi m_{h_1}}\sqrt{1-\dfrac{4
    m_{A_1}^2}{m_{h_1}^2}}\,.
\label{hd31}
\end{equation}

To compare the partial width of the non-standard Higgs decay of the
SM-like Higgs state, Eq.~(\ref{hd31}), with the Higgs
decay rates into the SM particles, we specify a set of benchmark
points (see Table~\ref{tab:benchmarks}).  For each benchmark scenario
a code that is automatically generated by {\tt FlexibleSUSY}
\cite{Athron:2014yba}
is used\footnote{We used an adapted version of {\tt FlexibleSUSY}-1.0.2
  which contains updates that will appear in the new version
{\tt FlexibleSUSY}-1.0.3. The generated (and modified) code can be supplied on
  request.} (based on {\tt SARAH}
  \cite{Staub:2010ty,Staub:2009bi,Staub:2010jh,Staub:2012pb,Staub:2013tta}
  and {\tt SOFTSUSY} \cite{Allanach:2001kg,Allanach:2013kza}) to determine
  the spectrum of the masses.  The complete one-loop self energies are
  included in the determination of all masses in the model, and
  leading two-loop contributions to the CP-even and CP-odd Higgs
  bosons from the NMSSM ($\oatas$ and $\oabas$) \cite{Degrassi:2009yq}
  and the MSSM ($\oatq$, $\oabatau$, $\oabq$, $\oatauq$ and $\oatab$)
  \cite{Degrassi:2001yf,Brignole:2001jy,Dedes:2002dy,Brignole:2002bz,Dedes:2003km}
  are included by using files provided by Pietro Slavich. The
  additional corrections that may arise due to our model are expected
  to be small, as either the new particles do not couple directly to
  the involved particles or their contributions are small due to
  suppressed couplings and/or large masses.

The couplings and branching ratios of the lightest CP-even Higgs state
were also obtained by calling routines generated by {\tt FlexibleSUSY} in a
small extension of the automatically generated code from {\tt FlexibleSUSY}.
{\tt FlexibleSUSY} uses {\tt SARAH-4.2.1} to derive analytical expressions, which
is independent of the derivation used to obtain the expressions
presented here. Therefore we were able to do an independent check of
the mass matrices presented in the previous section and of the
coupling $G_{h_1 A_1 A_1}$ given in Eq.~(\ref{hd30}) by comparing our code from
{\tt FlexibleSUSY} numerically against an alternative Mathematica code based
on those expressions.

Additionally, we cross-checked the thus obtained total width and
branching ratios for the lightest CP-even Higgs boson against the ones
obtained from the code {\tt HDECAY} \cite{hdecay}, respectively its
extension {\tt eHDECAY} \cite{ehdecay}. The Fortran program {\tt
  HDECAY}, originally written for the calculation of the decay widths
and branching ratios of the SM and the MSSM Higgs bosons, has been
extended to allow for the possiblity to change the couplings of the SM
Higgs boson by global modification factors. Using the modification
factors, i.e.~the ratios of the couplings of the lightest CP-even
Higgs boson in our model to the SM particles with respect to the
corresponding couplings of the SM Higgs boson of same mass, as inputs
in {\tt HDECAY}, we are able to compute the decay rates of $h_1$ into
SM particle final states. Implementing in addition the partial decay
width $h_1 \to A_1 A_1$ generated by {\tt FlexibleSUSY}, {\tt HDECAY}
can be used to also calculate all branching ratios.\footnote{Note,
  that also the decay $h_1 \to Z A_1$ is in principle possible. In all
  benchmarks scenarios, however, this decay is kinematically closed or
  strongly suppressed.} This procedure allows us to profit from the
state-of-the-art QCD corrections implemented in {\tt HDECAY}, which
can be taken over to our model.\footnote{The electroweak corrections
  cannot be taken over. They are consistently included only for the SM
  part of the decay widths, {\it cf.}~\cite{ehdecay}.} It should be
noted, that in the loop-induced couplings to gluons and photons,
respectively, the SUSY-related loops are not taken into account,
however, as the option of applying coupling modifications in {\tt
  HDECAY} only applies to the SM Higgs boson. As in our scenarios the
sfermion and charged Higgs boson masses are heavy, the change should
be marginal only. The branching ratios for the decay $h_1 \to A_1 A_1$
given for our benchmark scenarios in Table~2, are the ones obtained by
this procedure.

To simplify our analysis we set $B=\mu=\xi=0$ and $\Lambda=0$.
This does not change the physics we are investigating here
  and still leaves us with one crucial Peccei-Quinn violating coupling
  $\kappa$. We also fix $\sigma=0.1$ and $\tan\beta\simeq
10$. A large value of $\tan\beta$ allows us to
  maxime the tree-level mass value of the lightest CP-even Higgs
  boson, so that the experimental value of $\sim 125$~GeV can be
  obtained more easily. The small value for $\sigma$ leads to VEVs that
can be much heavier than the SUSY breaking scale, {\it cf.}~Eq.~(\ref{hd3}).
Then, in order to find an appropriate set of benchmark
points, we vary $\lambda$, $\kappa$, $A_{\kappa}$, $A_{\lambda}$,
$A_{\sigma}$, $A_t$, $m_{Q_3}^2$, $m_{u_3^c}^2$,
$\varphi$, $s$ and $\theta$. In all our benchmark
scenarios $m_{S}^2$ and $m_{H_u}^2$ are negative which
ensures that the Higgs fields $S$ and $H_u$ acquire VEVs that result in
non-zero VEVs for the other Higgs fields $H_d$, $S$ and
$\varphi$. This should trigger the breaking of the $SU(2)_L\times
U(1)_Y\times U(1)_{N}$ symmetry down to the $U(1)_{\rm em}$. The soft
scalar masses associated with the superpartners of the left-handed
and right-handed components of the top quark and the mixing in the stop
sector are chosen such that the SM-like Higgs state has a mass of
approximately $125-126\,\mbox{GeV}$.

To construct benchmark scenarios which are consistent with
cosmological observations, it is important to guarantee that they lead
to relic densities $\Omega_{\mathrm{CDM}} h^2$ that are not larger
than the result given by PLANCK \cite{Ade:2013zuv}:
\be
\Omega_{\mathrm{CDM}} h^2 = 0.1187 \pm 0.0017 \;.
\label{hd32}
\ee
A theory predicting a greater relic density than the PLANCK result is
basically ruled out, assuming standard pre-BBN cosmology. A theory
that predicts less dark matter cannot be ruled out in the same way, but
would require to have other contributions to the dark matter relic
density. Since the dark matter density is inversely proportional to
the annihilation cross section at the freeze-out temperature, this
cross section has to be sufficiently large. In the $E_6$ inspired SUSY
models considered here, the cold dark matter  density is formed by the
lightest neutralino states. At first glance the neutralino sector in
these models is more complicated than the one in the MSSM. Indeed, it
contains eight states which is twice as large as the one for the
MSSM. However an analysis of the corresponding neutralino mass matrix,
which is specified in the Appendix A, indicates that this matrix
given in the basis
  $(\tilde{H}^0_d,\,\tilde{H}^0_u,\,\tilde{W}_3,\,\tilde{B},\,\tilde{B}',\,\tilde{S}\cos\theta-\tilde{\overline{S}}\sin\theta,\,\tilde{S}\sin\theta+\tilde{\overline{S}}\cos\theta,\,\tilde{\phi})$
  has a rather simple structure. The
  $\tilde{H}^0_d,\,\tilde{H}^0_u,\,\tilde{W}_3,\,\tilde{B}$ denote the
  fermion components of the corresponding Higgs doublet and gauge
  fields, $\tilde{B}'$ the gaugino related to the $Z'$ vector
  superfield, and finally,
  $\tilde{S},\tilde{\overline{S}},\tilde{\phi}$ the fermion components
  of the SM singlet Higgs superfields $S,\overline{S}$ and $\phi$.
From Eq.~(\ref{a4}) it follows that the masses of two
neutralino states, that are linear superpositions of $\tilde{B}'$ and
$\tilde{S}\cos\theta - \tilde{\overline{S}}\sin\theta$, are controlled
by the $Z'$ boson mass.
In the limit where $\lambda$ is small and
$M_{Z'}\gg M_S$ these  states decouple from the rest of the
sparticle spectrum. As can be seen from Eq.~(\ref{a4}), two other neutralino
eigenstates are formed by the linear superpositions of
$\tilde{S}\sin\theta+\tilde{\overline{S}}\cos\theta$ and
$\tilde{\phi}$. In the
situation where $\kappa$ is much smaller than $\lambda$ and $\sigma$,
the masses of these eigenstates are approximately
\be
m_{\chi^0_{5,6}}\simeq  \dfrac{\sigma\varphi}{2\sqrt{2}}\left(1 \pm \sqrt{1+4\,\dfrac{s^2}{\varphi^2}}\right)\sim M_S\,.
\label{hd33}
\ee
When $\lambda$ is small and $M_S\gg M_Z$  the mixing of these states
with the MSSM-like neutralino (the superposition of $\tilde{H}^0_d$,\,
$\tilde{H}^0_u$,\,$\tilde{W}_3$ and $\tilde{B}$) is also strongly suppressed.
Thus, if the corresponding neutralino states are not the lightest ones,
then they can be ignored in first approximation. This permits us
to reduce the $8\times 8$ matrix, Eqs.~(\ref{a2})--(\ref{a5}), to a
$4\times 4$ mass matrix which is rather similar to the one in the
MSSM. In the limit  $M_S\gg M_Z$ the masses of the MSSM-like
neutralino states, which are predominantly bino, wino and higgsino,
are set by $M_1$, $M_2$ and $\lambda s \cos\theta/\sqrt{2}$,
respectively.

The qualitative pattern of the neutralino spectrum discussed above
reveals that for sufficiently small values of $\lambda$ the lightest
neutralino tends to be a higgsino dominated state. If this
higgsino-like state is lighter than $1\,\mbox{TeV}$, then it leads to
a cold dark matter density that is less than the observed value
\cite{ArkaniHamed:2006mb}. Therefore, in all benchmark scenarios
specified in Table~\ref{tab:benchmarks}, the value of the coupling
$\lambda$ is chosen
such that the lightest neutralino is predominantly higgsino with mass
below $1\,\mbox{TeV}$.

In summary, we choose the following values for our benchmark
scenarios:
\begin{eqnarray}
&& \hspace*{-1cm}
\underline{\mbox{The soft SUSY breaking left- and right-handed mass
    parameters: }} \nonumber \\
&& \begin{array}{lcl}
\hspace{-1cm}&& m_{Q_{1,2}}^2 = m_{u_{1,2}^c}^2 = m_{L_{1,2,3}}^2 =
m_{e_{1,2,3}^c}^2 = m_{L_4}^2 = m_{\overline{L}_4}^2 = 100 \mbox{ TeV}^2 \,, \\
&& m_D^2 = m_{\overline{D}}^2= m^2_{H^d_{1,2}} = m^2_{H^u_{1,2}} = 4 \mbox{ TeV}^2
\end{array} \label{eq:parameters1} \\[0.2cm]
&& \hspace*{-1cm} \underline{\mbox{The soft SUSY breaking gaugino mass parameters:
  }} \nonumber \\
&& \quad M_1 = 600 \mbox{ GeV} \;, \quad M_2 = 1.2 \mbox{ TeV}
\;, \quad M_3 = 3.6 \mbox{ TeV} \\[0.2cm]
&& \hspace*{-1cm} \underline{\mbox{The coupling values: }} \nonumber \\
&& \mu = B =\xi = 0, \; \;  \Lambda = 0, \; \\
&&\kappa_{ij} = 0.5 \,\delta_{ij}  \\
&& f_{i\alpha} = \tilde{f}_{i\alpha} = g^D_{ij} = h^E_{ij} =0 \,\\
&& \tilde{\sigma} = 0, \;\mu_L = 10 \mbox{ TeV}   \\
&& \hspace*{-1cm} \underline{\mbox{The mixing angle: }}
\tan\beta^{\overline{\text{DR}}}(M_Z) = 10 \;,  \label{eq:parameters7}
\end{eqnarray}
where the masses in Eq.~(37) are associated with the soft scalar
masses of the scalar components of superfields listed in Eq.~(7) and
$\delta_{ij}$ in Eq.~(40) denotes the Kronecker $\delta$.
All other values are specified in Table~\ref{tab:benchmarks}, with the
exception of those trilinear couplings that are zero.
The parameters are chosen such that $m_h \approx 125$ GeV. The SM
parameters are chosen as
\begin{eqnarray}
\begin{array}{lcllcl}
\alpha_{\text{em}}(M_Z) &=& 1/127.916
\,, & \, \alpha_s(M_Z) &=&  0.1184 \;, \\[0.1cm]
 M_Z &=& 91.1876 \mbox{ GeV} \,, & \, M_W &=& 80.404 \mbox{ GeV} \;,
 \\
m_t &=& 173.18 \mbox{ GeV} \,, & \,
m_b(m_b)^{\overline{\text{MS}}} &=& 4.2 \mbox{ GeV} \,, \quad m_\tau =
1.777 \;.
\mbox{ GeV}
\end{array}
\end{eqnarray}

As in all benchmark scenarios we took care to choose $m_{A_1}$ small
enough that the decays of the SM-like Higgs state into a pair of the
lightest pseudoscalar Higgs bosons are kinematically allowed, it is
important to ensure that the lightest CP-odd Higgs boson could not
have been detected in the collider experiments to date. In this
context it is worth noting that in the case of a hierarchical
structure of the Higgs spectrum, which is caused by a large SUSY
breaking scale, the couplings of this pseudoscalar state to the SM particles are
naturally suppressed. Indeed, as was pointed out in subsection 3.2,
the lightest CP-odd Higgs boson is predominantly a superposition of
the imaginary parts of the SM singlet fields $\overline{S}$, $S$ and
$\phi$. These do not couple directly to the SM
  particles and, furthermore, the mixing between these singlet fields
and the neutral components of the Higgs doublets is rather small in
this case. As a consequence in all benchmark scenarios presented in
Tables~\ref{tab:benchmarks} and \ref{tab:bench2} the absolute value of
the relative coupling of the lightest
pseudoscalar Higgs to the $Z$-boson and the SM-like Higgs state, $R_{Z
  A_1 h_1}$, is always smaller than $10^{-3}-10^{-4}$. All other couplings of
the lightest CP-odd Higgs boson to the SM particles are also extremely
small. This state therefore has escaped detection in past and present
collider experiments.

Due to the hierarchical structure of the Higgs spectrum the lightest
SM-like Higgs state has couplings close to the SM values. Its coupling
values $R_{XXh_1}$ to the SM particles $X=V,f_u,f_d$ ($V\equiv W,Z$,
$f_u\equiv u,c,t$, $f_d\equiv d,s,b,e,\mu,\tau$) relative to the ones of the SM Higgs
boson, are given in Table~\ref{tab:bench2} for the various
benchmarks scenarios. The $h_1$ couplings deviate by at most 4\% from
the SM Higgs couplings. These coupling values feed into the
production processes and the branching ratios of the SM-like state
$h_1$. It has to be made sure
that the $\mu$-values, {\it i.e.}~the ratios of $h_1$ production cross
section times branching ratio normalized to the corresponding SM
values, agree within the respective errors with the $\mu$-values reported by
the LHC experiments for the various final states. For
definiteness we take here the values given in Refs.~\cite{atlasnotemu}
and \cite{cmsnotemu}. We follow the procedure of
Ref.~\cite{King:2012is} and combine the signal rates and errors of
the two experiments according to Eq.~(5) in \cite{Espinosa:2012vu}. We
require the $h_1$ $\mu$-values to be within 2 times the 1$\sigma$
interval around the respective best fit value.  The combined signal
rates and errors are given in Table~\ref{tab:combvalues}.
\begin{table}[!h]
  \centering
  \begin{tabular}{|c||c|c|}
    \hline
channel & best fit value & $2\times 1\sigma$ error \\ \hline
$VH \to Vbb$ & 0.97 & $\pm 1.06$ \\ \hline
$H\to \tau\tau$ & 1.02 & $\pm 0.7$
\\ \hline
$H\to \gamma\gamma$ & 1.14  & $\pm 0.4$
\\ \hline
$H \to WW$ & 0.78 & $\pm 0.34$
\\ \hline
$H\to ZZ$ & 1.11 & $\pm 0.46$
\\ \hline
\end{tabular}
\caption{The combined ATLAS and CMS signal rates with errors for the
  $bb,\tau\tau,\gamma\gamma,WW$ and $ZZ$ final states. Apart from the $bb$
  final state, where Higgs-strahlung $VH$ ($V=W,Z$) is the production
  channel, they are based on the inclusive production cross
  section. Details can be found in Refs.~\cite{atlasnotemu} and
  \cite{cmsnotemu}.}
\label{tab:combvalues}
\end{table}
For the calculation of the $\mu$-values we have assumed the dominant
production cross section to be given by gluon fusion. Subsequently, we
have approximated the ratios of the $h_1$ and the SM gluon fusion
production cross sections by the ratio of their decay widths $\Gamma_{gg}$ into a
gluon pair. The gluon decay widths have been calculated with {\tt HDECAY} in
both cases, as outlined above. The program includes the higher order QCD
corrections to this decay.\footnote{For details and a recent
  discussion, see {\it e.g.}~\cite{ehdecay,Baglio:2013iia}.} The
approximation of the production cross section ratios by the decay
width ratios is valid within about 10-20\% depending on the scenario
\cite{King:2014xwa}. The $\mu$-value for $h_1$ into the final state
$XX$ is hence given by
\begin{equation}
\mu_{XX} (h_1) \approx \frac{\Gamma_{gg} (h_1) BR(h_1 \to
  XX)}{\Gamma_{gg} (H^{SM}) BR(H^{SM} \to XX)} \;,
\end{equation}
where $H^{SM}$ denotes the SM Higgs boson with the same mass as
$h_1$. The branching ratios $BR$ again have been obtained with {\tt
  HDECAY}. For the $b\bar{b}$ final state, however, Higgs-strahlung
$Vh_1$ ($V=Z,W$) is the production channel to be used, in accordance with the
experiments. This cross section is given by the SM value multiplied
with the squared coupling ratio $R_{VVh_1}$ to gauge bosons. For the
$b\bar{b}$ final state we hence have
\begin{equation}
\mu_{b\bar{b}} (h_1) \approx \frac{\sigma_{Vh_1} BR(h_1 \to
  b\bar{b})}{\sigma_{VH^{SM}} BR(H^{SM} \to b\bar{b})} =
R_{VVh_1}^2 \frac{BR(h_1 \to b\bar{b})}{BR(H^{SM} \to b\bar{b})} \;.
\end{equation}

Let us now turn to the discussion of the five benchmark
  scenarios BMA--BME, summarized in Tables~\ref{tab:benchmarks} and
  \ref{tab:bench2}. Table~\ref{tab:benchmarks} lists the parameter
  values defining the scenarios (in addition to
  those values common to all benchmarks, given in
  Eqs.~(\ref{eq:parameters1})-(\ref{eq:parameters7})) and the
  corresponding scalar Higgs masses
  $m_{h_1}$--$m_{h_5}$ and the pseudoscalar ones $m_{A_1}$--$m_{A_3}$,
  with the SM-like Higgs boson given by the lightest scalar $h_1$. In
  Table~\ref{tab:bench2} we give for each benchmark scenario the coupling
  $G_{h_1 A_1 A_1}$, relevant for the non-standard decay $h_1 \to A_1
  A_1$, and the ratios $R_{XXh_1}$ of the couplings of the SM-like
  $h_1$ with respect to the ones of the SM Higgs boson for the couplings to
  a pair of massive gauge bosons $V$, of up-type quarks $f_u$ and of
  down-type quarks $f_d$. The Table contains the $\mu$-values in
  the LHC Higgs search final states and finally the partial decay width and the
  branching ratio for the non-standard decay of $h_1$ into a pair of
  lightest pseudoscalars, along with the $h_1$ total decay
  width. For all scenarios, the $\mu$-values are
    within 2 times the 1$\sigma$ error interval around the
    experimentally measured values.\footnote{The
      difference in the $\mu$-values for the $WW$ and $ZZ$ final
      states arises from different branching ratios. Although the
      coupling modifications are the same for these final states, the
      branching ratios differ, as in {\tt eHDECAY} electroweak
      corrections are included in the decay width in the term linear in the SM
      amplitude. For details, see \cite{ehdecay}.}
\begin{table}[h!]
\begin{center}
{\small
  \noindent \begin{tabular}{| c || c | c |c| c| c|}
\hline
                                  &    BMA                 &  BMB         &     BMC          &    BMD            &     BME        \\
\hline
$\lambda$    &   0.100 & 0.090  & 0.100   &    0.090   &   0.090  \\
$\sigma$      &   0.1  & 0.1  &  0.1 &  0.1  & 0.1  \\
$\kappa$      &    0.03  & 0.001   & 0.03  &    0.0273  & 0.001      \\
$A_{\lambda}$ [GeV]   & 600     & 2222   & 800   &  2222   &  4444   \\
$A_{\sigma}$ [GeV]    & 1200    & 1200   & 1400  &  1200   & 2400  \\
$A_{\kappa}$ [GeV]    & 1013    & 1000   & 1023  &  1026   & 2200  \\
$A_{t}$ [GeV]        & -1186   & -1171  &-5944  & -1170   & -1163\\
\hline
$m_{Q_3}^2$ [GeV]$^2$     &  3.0 $\cdot 10^6$ & 1.0 $\cdot 10^6$ & 3.0
$\cdot 10^6$ & 2.0 $\cdot 10^6$ & 1.0 $\cdot 10^6$ \\
$m_{u^c_3}^2$ [GeV]$^2$  &   2.0 $\cdot 10^8$ & 1.0 $\cdot 10^8$ & 2.0
$\cdot 10^8$ & 8.0 $\cdot 10^7$ & 8.0 $\cdot 10^7$ \\
\hline
$\varphi$  [TeV]  &    6  &  6  & 	6   &   6   & 12  \\
$s$ [TeV]   &  8  & 8  & 8 &   8 &  16 \\
$\tan\theta$ &  0.91 &  0.91 & 0.91 &   0.91 & 0.91 \\
\hline
$M_1 = M_1^\prime$ [GeV] &    600  &  600  & 600  &  600 & 1200 \\
$m_{\chi_1^0}$ [GeV]  &    420 & 376 & 419 & 377 &   761\\
\hline
$m_S^2$ [GeV]$^2$ &  -7.023 $\cdot 10^5$ & -1.51 $\cdot 10^6$  &
-6.287 $\cdot 10^5$ &  -1.135 $\cdot 10^6$        & -2.282 $\cdot 10^6$      \\
$m_{\overline{S}}^2$ [GeV]$^2$  &    1.303 $\cdot 10^6$  &   1.918
$\cdot 10^6$  & 	1.400 $\cdot 10^6$ &  1.669 $\cdot 10^6$  &
3.655 $\cdot 10^6$        \\
$m_\phi^2$ [GeV]$^2$ &   6.145 $\cdot 10^4$  &  1.292 $\cdot 10^5$  &
1.353 $\cdot 10^5$ & 7.045 $\cdot 10^4$  & 5.219 $\cdot 10^5$  \\
$m_{H_d}^2$ [GeV]$^2$  &   8.267 $\cdot 10^5$ & 7.132 $\cdot 10^6$ & 1.306
$\cdot 10^6$     &  7.023 $\cdot 10^6$ & 2.684 $\cdot 10^7$ \\
$m_{H_u}^2$ [GeV]$^2$  &   -2.419 $\cdot 10^6$ & -1.063
$\cdot 10^6$
& -2.977 $\cdot 10^6$ & -2.448 $\cdot 10^5$   & -1.597 $\cdot 10^5$  \\
\hline
$m_{Z^\prime}$ [TeV]  &  2.956  & 2.964 & 2.956  & 2.961 &  5.939 \\
$m_{H^\pm}$ [GeV] &   799  & 2550 & 1057  & 2550 & 5123 \\
$m_{A_1}$ [GeV]  &	35.37 & 33.01 & 51.53 &  18.97 & 28.41 \\
$m_{A_2}$ [GeV]   &	791 & 1206 & 1051 &  1168 & 2430 \\
$m_{A_3}$ [GeV] &    1159 & 2547 & 1257 & 2548  &   5122 \\
$m_{h_1}$ [GeV] & 126.156 & 	125.76  & 125.881 & 125.699  & 126.225 \\
$m_{h_2}$ [GeV]  & 387  & 460 & 267  & 393  & 852 \\
$m_{h_3}$ [GeV]  & 791  & 795 & 1016 & 925  & 1569 \\
$m_{h_4}$ [GeV]  & 936 & 2547  & 1051  & 2548 & 5122  \\
$m_{h_5}$ [GeV]  & 3099 & 3093  & 3125  & 3104 & 6188 \\
\hline
\end{tabular} }
\end{center}
\caption{Parameters defining the benchmark points BMA--BME, with the associated
  Higgs masses.}
\label{tab:benchmarks}
\end{table}
  As can be inferred from Table~\ref{tab:bench2}, the
  branching ratios of the non-standard Higgs decays $h_1$ into $A_1
  A_1$ vary from $10^{-3}\,\%$ to $21\%$ for the various
  benchmarks. The table shows, that their size depends rather strongly
  on the absolute value of the coupling $\kappa$. Reasonably large
  branching ratios ($\gsim 1\%$) of these
Higgs decays can be obtained for $\kappa$ values of ${\cal O}(0.01)$. Thus we
obtain in benchmark scenario BMA with $\kappa=0.03$ a branching ratio
of $\sim 8$\%. In scenario BMC, where also $\kappa= 0.03$, the
parameters are further optimized to get a large branching ratio,
resulting in $BR(h_1 \to A_1 A_1) \approx 0.21$. The scenario BMD has
$\kappa =0.0273$ and a non-standard branching ratio of 0.017. The
branching ratio is smaller here, as by increasing $A_\lambda$,
resulting in a larger charged Higgs mass, we are getting closer to the
SM-limit, as can be seen from the coupling ratios to the SM
particles, which are almost one.
\begin{table}[t!]
\begin{center}
{\small
  \noindent \begin{tabular}{| c || c | c |c| c| c|}
\hline
&    BMA  &  BMB &     BMC  &  BMD   & BME \\
\hline
$R_{f_u f_u h_1}$  & -0.9974  & -0.9997 & 0.9888  & 0.9994
& -0.9999 \\
$R_{f_d f_d h_1}$ & -1.0412 & -1.0036 & 1.0142 & 1.0032 & -1.0013\\
$R_{VVh_1}$ & -0.99789  & -0.9997 & 0.9891 & 0.9995 & -0.9999 \\
$R_{ZA_1h_1}$ & -1.608 $\cdot 10^{-4}$ & 1.626 $\cdot 10^{-4}$ &
6.634 $\cdot 10^{-7}$ & -1.082 $\cdot 10^{-4}$ & 8.142 $\cdot 10^{-5}$\\
$G_{h_1 A_1 A_1}$ [GeV]  &  -1.2704  &  -0.0270  & 2.5782  &
0.5136  &  -0.0149\\
\hline
$\mu_{bb}$ & 0.9684 & 1.0317 & 0.7883  & 0.9841 & 1.0314 \\
$\mu_{\tau\tau}$ & 0.8731 & 0.9350 & 0.7865 & 0.9838 & 0.9347 \\
$\mu_{ZZ}$ & 0.7558 & 0.8741 & 0.7480 &  0.9765 & 0.8786 \\
$\mu_{WW}$ & 0.7247 & 0.8383 & 0.7483 & 0.9766 & 0.8422 \\
$\mu_{\gamma\gamma}$ & 0.8180 & 0.9461 & 0.7482 & 0.9766 & 0.9499 \\
\hline
$BR(h_1\rightarrow A_1A_1)$  &  0.0818  &  4.415 $\cdot 10^{-5}$
&  0.2078  & 0.0172 & 1.391 $\cdot 10^{-5}$ \\
$\Gamma(h_1\rightarrow A_1A_1)$ [GeV] & 4.215 $\cdot 10^{-4}$ &
1.967 $\cdot 10^{-7}$ & 1.206 $\cdot 10^{-3}$  &  7.960 $\cdot 10^{-5}$ &
6.270$\cdot 10^{-8}$\\
$\Gamma_{\text{tot}}$ [GeV]   &  5.154 $\cdot 10^{-3}$ &  4.456 $\cdot 10^{-3}$
 & 5.805$\cdot 10^{-3}$ & 4.618 $\cdot 10^{-3}$ & 4.451 $\cdot 10^{-3}$\\
\hline
\end{tabular} }
\end{center}
\caption{For the benchmarks BMA--BME, the couplings
  and coupling ratios of the SM-like scalar Higgs $h_1$ as well as the
  $\mu$-values for the LHC Higgs search channels. The
    ratio $R_{ZA_1 h_1}$ denotes the ratio of the coupling in the SUSY
    model under consideration with respect
  to the corresponding coupling in the MSSM.
  The last three lines show the branching ratio and partial width for
  the decay $h_1 \to A_1 A_1$ and the total width.}
\label{tab:bench2}
\end{table}
Therefore, the non-standard coupling
$G_{h_1 A_1 A_1}$ is smaller compared to BMA and BMC and hence also
the corresponding branching ratio. Both in scenario BMB and BME the
$\kappa$ value is chosen to be small, $\kappa = 0.001$. This results in
$BR(h_1 \to A_1 A_1) = 4.4 \cdot 10^{-5}$ in scenario BMB. In BME the
VEV $s$ has been increased and hence the mass of the $Z'$ which is
$M_{Z'} \approx 6$~TeV here, resulting in an even smaller branching
ratio. With $h_1$ non-standard branching ratios of ${\cal O}(10^{-5})$ and
$\mu$ rates that are compatible with the SM, this scenario will not be
quickly ruled out by the LHC.\footnote{The lightest
    CP-odd Higgs state that originates from decays
  of the SM-like Higgs boson, predominantly decays into either a pair
  of $b$-quarks or $\tau$-leptons giving rise to four fermion final
  states.}

\begin{figure}[t!]
\begin{center}
\includegraphics[width=0.8\textwidth]{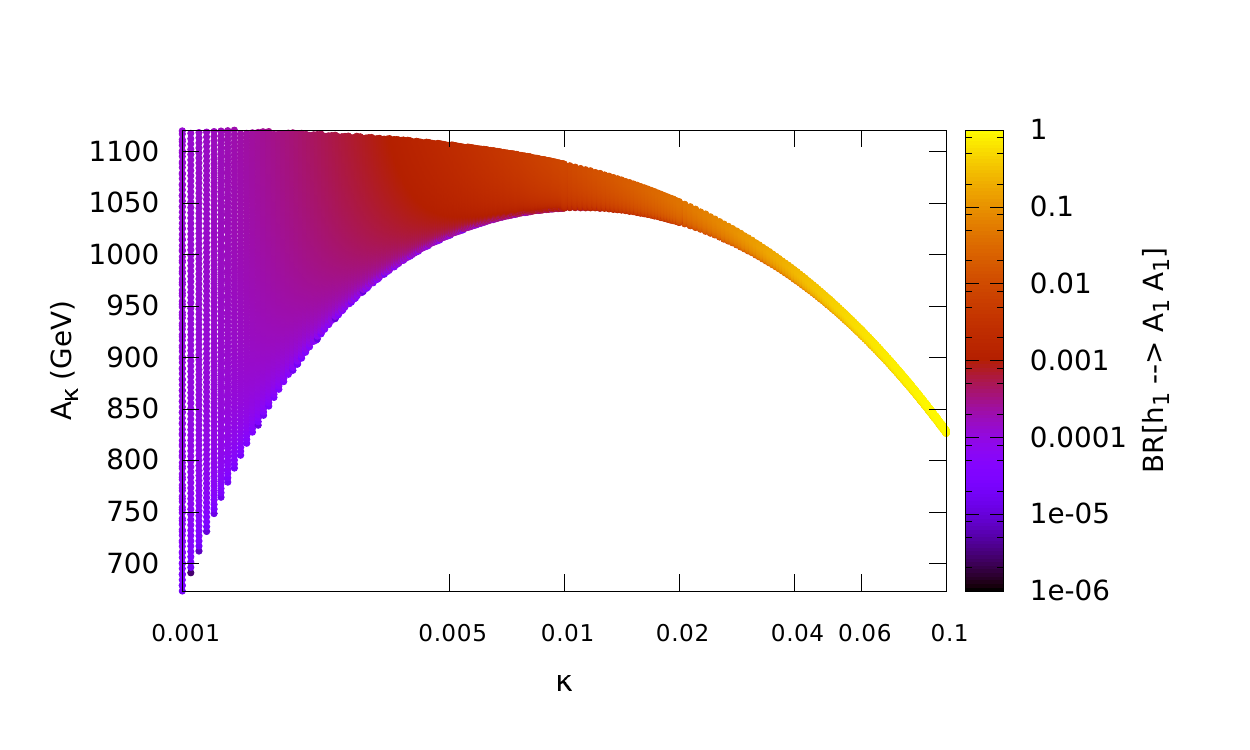}
\caption{Colour contours of the branching ratio for $h_1\rightarrow
  A_1 A_1$ in the $\kappa$--$A_\kappa$ plane. An adapted version of
  {\tt FlexibleSUSY} was used to calculate the mass spectra and
  the branching ratios.  All other parameters are fixed to the values
  of BMA.  For each value of $\kappa$ there is a lower limit on
  $A_\kappa$ where the the $BR$ is zero because the pseudoscalar is too
  heavy and an upper limit above which there is a pseudoscalar
  tachyon. \label{fig:tuning}}
\end{center}
\end{figure}
For the non-standard decays to take place, the pseudoscalar mass must
be small enough. As follows from Eqs.~(\ref{hd21}) such a small
$m_{A_1}$ can always be obtained by tuning the parameter
$A_{\kappa}$. The degree of tuning is illustrated in
Fig.~\ref{fig:tuning}, where the region of the parameter space that
leads to a sufficiently small $m_{A_1}$ for non-standard Higgs decays,
is shown in the $\kappa$--$A_{\kappa}$ plane.  One can see that with
decreasing $\kappa$, the range of $A_{\kappa}$ that results in a
sufficiently small $m_{A_1}$ becomes considerably wider.  This
corresponds to a smaller degree of fine-tuning required to obtain a
sufficiently light pseudoscalar.\footnote{The branching ratios for
  this figure have been estimated using generated routines from {\tt
    FlexibleSUSY}, which allows for a fast scan over
  the parameter range. We have checked for a few points that the
  branching ratios from this simple estimate agree with the ones from
  {\tt HDECAY} well enough for our purposes. The deviations are due to
  approximation made for the $h_1$ decays into SM particles, which in
  our simple estimate were set equal to the ones of a SM Higgs boson
  with same mass. Further differences arise due to the inclusion of
  higher order corrections in {\tt HDECAY}.}  Therefore, in order to
get a pseudoscalar with mass around $40-60\,\mbox{GeV}$ for the
$\kappa$ values of ${\cal O}(0.03)$ of BMA, BMC and BMD, a fine-tuning
of order $1\%$ is needed.  It turns out that BME is strongly
fine-tuned ($\sim 1$\%) as well. This is because the SUSY breaking
scale is doubled compared to the other scenarios and the value of
$A_\kappa$ is twice the value of $A_\kappa$ in benchmark BMB. For
scenario BMB, on the other hand, the fine-tuning is $\lsim 10$\%.

Both Fig.~\ref{fig:tuning} and
Tables~\ref{tab:benchmarks} and \ref{tab:bench2} demonstrate
that the branching ratios of the decays of the SM-like Higgs boson
into a pair of pseudoscalar Higgs states become smaller when $\kappa$
decreases.  This is not a surprising result.
Indeed, in the PQ symmetric limit, when the global PQ symmetry is only broken
spontaneously, the coupling $G_{h_1 A_1 A_1}$ is set by (see for example \cite{Huang:2014cla})
\begin{equation}
G_{h_1 A_1 A_1}\simeq \dfrac{m_{h_1}^2}{2M_{PQ}}\,\varepsilon\,,
\label{hd000}
\end{equation}
where $M_{PQ}$ is the PQ symmetry breaking scale and $\varepsilon$
represents a suppression associated with the mixing between the SM-like Higgs
state and heavy CP-even Higgs states that induce the breakdown of the PQ symmetry.
Equation~(\ref{hd000}) also determines the size of $G_{h_1 A_1 A_1}$ when
the PQ violating couplings are rather small and $m_{A_1}$ is naturally very light.
A simple estimate using the values of the VEVs of the SM singlet fields given in
Table~\ref{tab:benchmarks} indicates that the absolute value of $G_{h_1 A_1 A_1}$
is expected to be considerably smaller than $1\,\mbox{GeV}$ when $\kappa\to 0$.
Thus one can expect that small values of $\kappa$ lead to small branching ratios
of the non-standard Higgs decays. When $\kappa\sim 0.001$ a pseudoscalar
state with mass around $40-60\,\mbox{GeV}$ can be obtained with very little
fine-tuning. However in this case, the branching ratios of the
non-standard Higgs decays are of the order of $10^{-4}$ and below.
At the same time when the PQ violating parameters are sufficiently
large so that one should fine-tune $m_{A_1}$ for $h_1 \rightarrow A_1 A_1$
to be kinematically allowed then there can be additional contributions to $G_{h_1 A_1 A_1}$
from the explicit PQ violating terms which can make its absolute value even larger than
$m_{h_1}^2 / (2 M_{PQ})$.  This is why we can obtain a large partial
width for this decay when $\kappa\gtrsim 0.01$.

We can hence summarize that in the model, that we introduced in order
to reduce the fine-tuning, from which simple $U(1)$-extended SUSY
models suffer, the non-standard decay of the SM-like Higgs state $h_1$
into a pair of pseudoscalars is possible.
Small $\kappa$ values lead naturally to light
pseudoscalar masses without fine-tuning. In this case, however, the
non-standard branching ratios are tiny, and it will be difficult to
test these exotic decays at the LHC.
For larger absolute values of $\kappa$, the corresponding branching ratio can
reach the level of a few per cent up to ${\cal O}(20\%)$. This can be
achieved, however, only at the price of fine-tuning, in order to get a
low enough pseudoscalar mass in this case for the decay to be
kinematically allowed.

\section{Conclusions}

In this paper we considered the non-standard decays of the SM-like
Higgs state within $U(1)$ extensions of the MSSM in which the extra
$U(1)$ gauge symmetry is broken by two VEVs of the SM singlet fields
$S$ and $\overline{S}$ with opposite $U(1)$ charges. Because in these
models $S$ and $\overline{S}$ can acquire very large VEVs along the
$D$-flat direction, the $Z'$-boson can naturally be substantially heavier
than the SUSY particles. This allows us to satisfy experimental
constraints and to alleviate the fine-tuning associated with the $Z'$-boson.
Such $U(1)$ extensions of the MSSM can possess an approximate
global $U(1)$ symmetry that gets spontaneously broken by the VEVs of $S$
and $\overline{S}$ leading to a pseudo-Goldstone boson in the particle spectrum.
If this pseudo-Goldstone state is considerably lighter than the lightest
CP-even Higgs boson then it may give rise to the decays of the SM-like Higgs
particle into a pair of pseudoscalars.

Here we studied such decays within well motivated SUSY extensions of
the SM based on the $SU(3)_C\times SU(2)_L\times U(1)_Y\times
U(1)_{N}\times Z_{2}^{M}$ group which is a subgroup of $E_6$. The
low-energy matter content of these $E_6$ inspired models includes
three $27$ representations of $E_6$, a pair of $SU(2)_L$ doublets
$L_4$ and $\overline{L}_4$, a pair of the SM singlets $S$ and
$\overline{S}$ with opposite $U(1)_{N}$ charges, as well as an SM
singlet $\phi$ that does not participate in the gauge interactions. To
suppress flavor changing processes at tree-level and to forbid the
most dangerous baryon and lepton number violating operators an extra
$\tilde{Z}^{H}_2$ discrete symmetry is imposed. We analysed the
spectrum of the CP-even and CP-odd Higgs bosons in these $E_6$
inspired models assuming that all SUSY breaking parameters are of the
order of the TeV scale. Expanding the Higgs potential we obtained
an analytical expression for the coupling of the lightest CP-even Higgs
state $h_1$ to a pair of the lightest Higgs pseudoscalars $A_1$. The dependence of
the branching ratio of the non-standard Higgs decay, $h_1 \to A_1 A_1$,
on the parameters in these SUSY models was examined. For simplicity,
we assumed that there is only one dimensionless coupling $\kappa$ in
the superpotential that explicitly violates the global $U(1)$
symmetry. When $\kappa$ vanishes the global $U(1)$ symmetry is
restored.

In order to illustrate the results of our analysis we specified a set
of benchmark scenarios with the SM-like Higgs mass around
$125-126\,\mbox{GeV}$. To ensure that the obtained benchmark points
are consistent with the measured value of the cold dark matter density
we chose the parameters such that the lightest neutralino is mainly
higgsino with a mass below $1\,\mbox{TeV}$. In this case the dark matter
density tends to be smaller than its observed value. The results of
our analysis indicate that the couplings of the lightest pseudoscalar
Higgs to the SM particles are always quite small. As a consequence,
although this pseudoscalar state can be rather light, it could escape
detection at former and present collider experiments.

We argued that the branching ratio of the decays of the lightest
CP-even Higgs boson into a pair of the lightest pseudoscalars depends
rather strongly on the absolute value of the coupling $\kappa$.
For absolute values of $\kappa$ which are substantially smaller
than $0.01$ the branching ratio of the non-standard Higgs decay in a
light pseudoscalar pair decreases considerably. 
Indeed, in the limit when $\kappa$ goes to zero the global
$U(1)$ symmetry is spontaneously broken and the coupling of
the lightest Higgs pseudoscalar to the SM-like Higgs becomes
extremely suppressed. As a result the branching ratio of the
non-standard Higgs decay tends to be negligibly small.
Therefore although for $\kappa\sim 0.001$
the lightest Higgs pseudoscalar with a mass of $40-60\,\mbox{GeV}$
can be obtained without fine-tuning the branching ratio of the SM-like
Higgs decays into a pair of the lightest CP-odd states is smaller than
$10^{-4}$. Decays with such small branching ratios will
be difficult to be tested at the LHC.

When $\kappa\gtrsim 0.01$ the branching ratio of the non-standard Higgs
decays can be larger than $1\%$. Nonetheless a fine tuning of at least
$1\%$ is required in this case to obtain a lightest pseudoscalar state
with mass of $40-60\,\mbox{GeV}$.  After being produced from the decay
of the lightest CP-even Higgs boson, the lightest CP-odd Higgs states
sequentially decay into a pair of either $b$-quarks or
$\tau$-leptons. Thus these decays of the lightest CP-even Higgs
boson result in four fermion final states.

We have found that with a modest fine tuning of
$A_{\kappa}$ one can obtain scenarios with
$h_1 \to A_1 A_1$ branching ratios of $\simeq 1\%$ and
acceptable values of dark matter relic density.

\vspace{-5mm}
\section*{Acknowledgements}
\vspace{-3mm}
We are grateful to S.~F.~King and A.~W.~Thomas for  fruitful discussions.
RN also thanks E.~Boos, V.~Novikov and M.~Vysotsky for useful comments
and remarks.
M.M. is supported by the DFG SFB/TR9 "Computational Particle Physics".
This work was supported by the University of Adelaide and the
Australian Research Council through the ARC Center of Excellence in
Particle Physics at the Terascale.


\newpage
\begin{appendix}
\section{The Neutralino Mass Matrix}
\setcounter{equation}{0}
\def\theequation{A.\arabic{equation}}

After the breaking of the gauge symmetry in the $E_6$ inspired SUSY
models under consideration all superpartners of the gauge and Higgs
bosons get non-zero masses. Because the extra vector superfield associated
with the $Z'$ boson and the extra SM singlet Higgs superfields $S$,
$\overline{S}$ and $\phi$ are electromagnetically neutral they do
not contribute any extra particles to the chargino spectrum. As a
consequence the chargino mass matrix and its eigenvalues remain almost
the same as in the MSSM, {\it i.e.}
\be
\ba{rcl}
m^2_{\chi^{\pm}_{1,\,2}}&=&\ds\frac{1}{2}\biggl[M_2^2+\mu_{\text{eff}}^2+2
M^2_{W}\pm\biggl.\\[2mm]
&&\biggr.\qquad\qquad\qquad\sqrt{(M_2^2+\mu^2_{eff}+2M^2_{W})^2-4(M_2\mu_{eff}-M^2_{W}\sin 2\beta)^2}
\biggr]\,,
\ea
\label{a1}
\ee
where $M_2$ is the $SU(2)_L$ gaugino mass and
\begin{equation}
\mu_{\text{eff}}=\ds\frac{\lambda s \cos\theta}{\sqrt{2}} \;.
\end{equation}
The non-observation of the lightest chargino at the collider experiments
implies that $|M_2|$, $|\mu_{\text{eff}}|\gtrsim 100\,\mbox{GeV}$.

The neutralino sector of the SUSY models under consideration involves
four extra neutralinos besides the four MSSM ones. One of them,
$\tilde{B}'$, is an extra gaugino coming from the $Z'$ vector
superfield. Three other states are the fermion components $\tilde{S}$,
$\tilde{\overline{S}}$ and $\tilde{\phi}$ of the SM singlet Higgs
superfields $S$, $\overline{S}$ and $\phi$. In the basis
$(\tilde{H}^0_d,\,\tilde{H}^0_u,\,\tilde{W}_3,\,\tilde{B},\,\tilde{B}',\,\tilde{S}\cos\theta-\tilde{\overline{S}}\sin\theta,\,\tilde{S}\sin\theta+\tilde{\overline{S}}\cos\theta,\,\tilde{\phi})$
the neutralino mass matrix can be written as
\be
M_{\tilde{\chi}^0}=
\left(
\ba{cc}
A & C^{T}\\[2mm]
C & B
\ea
\right)\,,
\label{a2}
\ee
with
\be
A=
\left(
\ba{cccc}
0 & -\dfrac{\lambda s \cos\theta}{\sqrt{2}} & \dfrac{gv}{2}\cos\beta & -\dfrac{g'v}{2}\cos\beta \\[3mm]
-\dfrac{\lambda s \cos\theta}{\sqrt{2}} & 0 & -\dfrac{gv}{2}\sin\beta & \dfrac{g'v}{2}\sin\beta \\[3mm]
\dfrac{gv}{2}\cos\beta & -\dfrac{gv}{2}\sin\beta & M_2 & 0\\[3mm]
-\dfrac{g'v}{2}\cos\beta & \dfrac{g'v}{2}\sin\beta & 0 & M_1
\ea
\right)\,,
\label{a3}
\ee
\vspace{3mm}
\be
B=
\left(
\ba{cccc}
M'_1 & g'_1 \tilde{Q}_S s & 0 & 0 \\[3mm]
g'_1 \tilde{Q}_S s & \dfrac{\sigma\varphi}{\sqrt{2}}\sin 2\theta & -\dfrac{\sigma\varphi}{\sqrt{2}}\cos 2\theta & 0 \\[3mm]
0 & -\dfrac{\sigma\varphi}{\sqrt{2}}\cos 2\theta & -\dfrac{\sigma\varphi}{\sqrt{2}}\sin 2\theta & -\dfrac{\sigma s}{\sqrt{2}}\\[3mm]
0 & 0 &  -\dfrac{\sigma s}{\sqrt{2}} & \sqrt{2}\kappa\varphi +\mu
\ea
\right)\,,
\label{a4}
\ee
\vspace{3mm}
\be
C=
\left(
\ba{cccc}
\tilde{Q}_{H_d} g'_1 v \cos\beta & \tilde{Q}_{H_u} g'_1 v \sin\beta & 0 & 0 \\[3mm]
-\dfrac{\lambda v}{\sqrt{2}}\sin\beta\cos\theta & -\dfrac{\lambda v}{\sqrt{2}}\cos\beta\cos\theta & 0 & 0 \\[3mm]
-\dfrac{\lambda v}{\sqrt{2}}\sin\beta\sin\theta & -\dfrac{\lambda v}{\sqrt{2}}\cos\beta\sin\theta & 0 & 0\\[3mm]
0 & 0 &  0 & 0
\ea
\right)\,,
\label{a5}
\ee where $M_1$, $M_2$ and $M'_1$ are the soft SUSY breaking gaugino masses for
$\tilde{B}$, $\tilde{W}_3$ and $\tilde{B}'$, respectively. Here we
neglect the Abelian gaugino mass mixing $M_{11}$ between $\tilde{B}$
and $\tilde{B}'$.  The top-left $4\times 4$ block of the mass matrix
in Eq.~(\ref{a2}) contains the neutralino mass matrix of the MSSM
where the parameter $\mu$ is replaced by $\mu_{\text{eff}}$. The lower right
$4\times 4$ submatrix represents extra neutralino states in this SUSY
model.

\end{appendix}


\end{document}